\documentclass[10pt]{bmc_article}

\usepackage{amsmath}
\usepackage{amsfonts}
\usepackage{amssymb}
\usepackage{latexsym}
\usepackage{indentfirst}
\usepackage{ifthen}  % Conditional 
\usepackage{multicol}   %Columns
\usepackage{cite} % Make references as [1-4], not [1,2,3,4]
\usepackage{url}  % Formatting web addresses  
\usepackage{graphicx,geometry}
\usepackage[dvips]{epsfig}
\usepackage{subfig}
\usepackage{lineno}

\def\VIZ#1{(\ref{#1})}      % use for references to formulae
\def\FISHER{\mathbf{F_{\mathcal{R}}}}
\def\FISHERR{\mathbf{F_{\mathcal{H}}}}
\def\RELENTR#1#2{\mathcal{H}\left({#1}\SEP{#2}\right)}
\def\RELENT#1#2{\mathcal{R}\left({#1}\SEP{#2}\right)}
\def\SEP{{\,|\,}}           % use for seperator in set defs, like

\def\LAWEXACT{Q_{[0,T]}^{\theta}}
\def\LAWAPPROX{Q_{[0,T]}^{\theta+\epsilon}}
\def\PATHS{\LAWEXACT}
\def\PATHSAPP{\LAWAPPROX}

\newboolean{publ}
\newenvironment{bmcformat}{\baselineskip10pt\sloppy\setboolean{publ}{false}}{\baselineskip10pt\sloppy}

\title{Parametric Sensitivity Analysis for Biochemical Reaction  Networks based on
Pathwise Information Theory}

\author{Yannis Pantazis$^1$\email{Yannis Pantazis - pantazis@math.umass.edu},
Markos A. Katsoulakis\correspondingauthor$^1$\email{Markos A. Katsoulakis\correspondingauthor - markos@math.umass.edu} and
Dionisios G. Vlachos$^2$\email{Dionisios G. Vlachos - vlachos@udel.edu}
}
      
\address{%
    \iid(1)Department of Mathematics and Statistics, University of Massachusetts, Amherst, MA, 01002. USA\\
    \iid(2)Department of Chemical Engineering, University of Delaware, Newark, Delaware, 19716. USA
}%

\date{\today}

\begin{document}
\begin{bmcformat}

\maketitle
\begin{abstract}

\noindent
{\bf Background:}
Stochastic modeling and simulation provide powerful predictive methods for the intrinsic
understanding of fundamental mechanisms in complex  biochemical  networks. Typically,
such mathematical models involve networks of coupled jump stochastic processes with
a large number of  parameters that need to be suitably calibrated against experimental
data. In this direction, the parameter sensitivity analysis of    reaction networks is an essential
 mathematical and computational  tool, yielding information regarding
the robustness and  the identifiability  of  model parameters. However, existing
sensitivity analysis approaches  such as  variants of the finite difference method  can have 
an overwhelming computational cost  in models with a high-dimensional parameter space.

\smallskip
\noindent
{\bf Results:}
We develop a sensitivity analysis methodology suitable for complex stochastic reaction networks
with  a large number of parameters. The proposed approach is based on Information Theory
methods and relies  on the quantification of information loss due to  parameter perturbations
between time-series distributions. For this reason, we need to work on path-space, i.e., the set
consisting  of all stochastic trajectories, hence the proposed approach is  referred to as ``pathwise".
The pathwise sensitivity analysis method is realized   by employing the rigorously-derived Relative
Entropy Rate (RER), which is  directly  computable from the propensity functions. A key aspect
of the method is that an  associated pathwise Fisher Information Matrix (FIM) is defined, which
in turn constitutes a gradient-free approach to quantifying parameter sensitivities. The structure
of the FIM  turns out to be block-diagonal,  revealing    hidden parameter dependencies and sensitivities
in reaction networks. 
%Furthermore, we suggest using not only exact stochastic simulation
%algorithms but also multi-scale numerical approximations
%of stochastic reaction networks (mean field, stochastic Langevin, $\tau$-leap, etc.) in order
%to derive efficient statistical estimators for the FIM.

\smallskip
\noindent
{\bf Conclusions:}
As a gradient-free method, the proposed sensitivity analysis provides a significant  advantage when
dealing with complex stochastic systems with a large number of parameters. In addition, the
knowledge of the structure of the FIM can allow to efficiently address questions on parameter
identifiability, estimation and robustness. The proposed method is tested and validated on
three biochemical systems, namely: (a) a protein production/degradation model
where explicit solutions are available, permitting a careful assessment of the method, (b) the p53 reaction network
where quasi-steady stochastic oscillations of the concentrations are observed, and for
which continuum approximations (e.g. mean field, stochastic Langevin, etc.) break down
due to persistent oscillations   between high and low populations, and (c) an Epidermal Growth
Factor Receptor  (EGFR) model  which is an example of a high-dimensional  stochastic reaction
network with more than 200 reactions and a corresponding number of parameters.
\end{abstract}


%\linenumbers

\ifthenelse{\boolean{publ}}{\begin{multicols}{2}}{}

\section{Background}
\label{intro}
The need of an intrinsic understanding of  the interplay between complexity and
robustness of biological processes and their corresponding design principles is
well-documented, see for instance \cite{Leibler:97, Doyle:02, Kitano:04, Donze:11,
Alon:12}. The concept of robustness  can be  described as  ``a property
that allows a system to maintain its functions against internal and external
perturbations" \cite{Kitano:04}. When referring to mathematical models of complex
biological processes, one of the mathematical tools to describe the robustness
of a system to perturbations is  sensitivity analysis which attempts to determine
which parameter directions (or their combinations)  are the most/least sensitive to
perturbations and uncertainty, or to errors resulting from experimental parameter estimation.
Recently there has been significant progress in developing sensitivity analysis
tools for low-dimensional stochastic processes, 
modeling well-mixed chemical reactions and biological networks. Some of the mathematical
tools included  log-likelihood methods and Girsanov transformations\cite{Glynn:90,Nakayama:94,Plyasunov:07}, 
polynomial chaos \cite{Kim:07},   finite difference methods and their variants 
 \cite{Rathinam:10, Anderson:12} and pathwise
sensitivity methods \cite{Khammash:12}. However, existing
sensitivity analysis approaches    can have 
an overwhelming computational cost,  either due to high variance in the gradient estimators,
or in models with a high-dimensional parameter space, \cite{Vlachos:12}.

The aforementioned methods focus on the sensitivity of stochastic trajectories and
corresponding averages. However, it is often the case that we are interested in
the sensitivity  of probability density functions (PDF), which are typically non-Gaussian
in nonlinear and/or discrete systems. In that latter direction, there is a
broad recent  literature relying on information theory tools,  and where  sensitivity is
estimated by using the  Relative Entropy and the   Fisher Information Matrix between
PDFs, providing a quantification  of information loss along different parameter
perturbations. We refer to \cite{Liu:06,Ludtke:08,Majda:10,Majda:11,Komorowski:11}
for the case when the  parametric PDF is explicitly known. For instance, in \cite{Majda:10}
the parametric PDF's structure is known  as it is obtained through an entropy
maximization subject to constraints. Knowing the form of the PDF allows to carry
out calculations such as estimating  the relative entropy and identifying  the most
sensitive parameter combinations. Furthermore, the pathwise PDFs are also known
in reaction networks when a Linear Noise Approximation
(LNA) is employed and for this case the relative entropy can  be explicitly computed
allowing thus to carry out parametric sensitivity analysis, \cite{Komorowski:11}. However, for
complex stochastic dynamics of large reaction networks, spatial Kinetic Monte Carlo
algorithms and molecular dynamics, such explicit formulas for the PDFs are  
not  available in general.

In \cite{Pantazis:Kats:13}, we address such challenges
by  introducing  a new   methodology for complex stochastic dynamics based on the
Relative Entropy Rate (RER) which  provides a measure of the sensitivity of the entire 
time-series distribution. Typically, the space of all such time-series is referred
in probability theory as the ``path space". RER measures the loss of information per
unit time in path space after an arbitrary perturbation of parameter combinations. RER
and the corresponding Fisher Information Matrix (FIM) become computationally feasible
as they admit explicit formulas which depend only on the propensity functions (see
\VIZ{RER:MP} and \VIZ{FIM:MP}, respectively).
In fact, we showed  in \cite{Pantazis:Kats:13} that  the proposed pathwise approach to
sensitivity analysis has   the following features: First, it is rigorously valid for the sensitivity
of  long-time, stationary  dynamics in path space, including for example bistable, periodic
and pulse-like  dynamics. Second, it is a gradient-free sensitivity analysis method
suitable for high-dimensional parameter spaces as the ones typically arising in complex
biochemical networks. Third, the RER method  does not require the explicit knowledge
of the equilibrium PDFs, relying only on   information for local dynamics and thus  making it
suitable for non-equilibrium steady state systems.
In \cite{Pantazis:Kats:13}, we demonstrated these features 
by focusing on  two classes of problems:   Langevin particle systems with either reversible
(gradient) or non-reversible (non-gradient) forcing,  highlighting the ability of the  method to carry
out sensitivity analysis in non-equilibrium systems; and spatially extended  Kinetic Monte Carlo
models,  showing that the method can  handle high-dimensional problems.

%%%% Novelty%%%%%%%%%%%%%%%%%
In this paper, we extend and apply the pathwise sensitivity analysis method in
\cite{Pantazis:Kats:13}  to biochemical reaction networks, and demonstrate
the intrinsic sensitivity structure of the network.
Such systems are  typically modeled as jump Markov processes and
they are simulated using either exact algorithms such as the Stochastic Simulation
Algorithm (SSA), \cite{Gillespie:77, Chatterjee:07, Slepoy:08} and the next-reaction
method \cite{Gibson:00},  or by employing approximations such as mean field ODEs,
tau-leap \cite{Gillespie:01} and stochastic Langevin methods \cite{Gillespie:00}.
%Our proposed method is based on the quantification of information loss along
%different model parameter perturbations between time-series distributions. This
%is achieved by employing rigorously derived  Relative Entropy Rate, which are
%directly  computable from the propensity functions only.
We show that the proposed pathwise method allows us  to discover the intrinsic
sensitivities of the reaction network by decomposing  the FIM  into diagonal blocks.
The block-diagonal structure of the proposed FIM reveals, in a straightforward way, the
sensitivity interdependencies between the  system parameters. For instance, if each
propensity function depends only on one parameter --usually the reaction constant-- then
the FIM is a diagonal matrix (see \VIZ{diagonal}). The sparse representation of the FIM
can be essential  in optimal experimental design as well as in parameter identifiability
and robustness where each subset of the parameters defined by a block of the FIM can
be treated separately.
Moreover, our earlier  rigorous analysis \cite{Pantazis:Kats:13} for the stationary regime
suggests suitable extensions in the transient case which are here  tested and validated. Finally,
we present strategies for efficiently and reliably  implementing the proposed method for 
high-dimensional, complex stochastic systems using an array of existing 
accelerated versions of the SSA algorithm such as mean field, stochastic Langevin,
$\tau$-leap approximations and their variants, \cite{Gillespie:00, Gillespie:01,
Rathinam:03, Chatterjee:05, Chatterjee:07}.

We test the proposed set of methods and computational strategies in three examples
of biochemical networks. First, we consider a prototypical protein production/degradation model,
i.e, a single-species birth/death model, with explicitly known formulas
for the stationary and the time-dependent distribution. This model serves
as a benchmark where the differences between the proposed pathwise FIM and the
stationary FIM are highlighted. Second, we study the parameter sensitivities of a p53
gene model for cell cycle regulation and response to DNA damage,   that incorporates
the feedback  between the tumor suppressor p53 gene and the oncogene Mdm2 \cite{Geva-Zatorsky:06}.
This is a reaction network that exhibits random oscillations in its steady state, and for
which continuum approximations of the SSA such as LNA break down due to persistent
oscillations between high and low populations. Using the proposed method, we also
study a far more complex network, the epidermal growth factor receptor (EGFR) model, 
describing  signaling processes between  mammalian cells \cite{Moghal:99, Hackel:99, Schoeberl:02}.
This is a high-dimensional system  both in the number of variables and parameters,
including 94 species and  207 reactions. Having a gradient-free method such as FIM
for this  example with parameter space of dimension 207 provides a significant advantage
over gradient methods such as finite differencing, where the computation of a very high number
of partial derivatives and/or directional derivatives is needed and with possibly significant variance
that scales with the dimension, \cite{Anderson:12}. By contrast, the eigenvalue/eigenvector
analysis of the proposed FIM identifies the order from  least to most  sensitive directions
(determined by the eigenvectors of the FIM) by the corresponding eigenvalues.

%%%%%%%%%%%%%ORGANIZATION%%%%%%%%%
In Methods, we present the derivation of the Relative Entropy Rate and its
corresponding Fisher Information Matrix for continuous-time jump Markov processes
as well as we reveal the block-diagonal structure of the FIM for commonly encountered
reaction networks, continued by the presentation of both unbiased and biased --but
accelerated-- statistical estimators for RER and FIM.
Then, in the Results, we apply and validate the proposed pathwise sensitivity
analysis methodology in three complex biological reaction networks.

\section*{Methods}
\label{methods:sec}
We consider a well-mixed reaction network with $N$ species, ${\bf S} = \{S_1,...,S_N\}$,
and $M$ reactions, ${\bf R} = \{R_1,...,R_M\}$. The state of the system at any time
$t\geq0$ is denoted by an $N$-dimensional vector ${\bf X}(t) = [X_1(t),...,X_N(t)]^T$ where
$X_i(t)$ is the number of molecules of species $S_i$ at time $t$. Let the $N$-dimensional
vector ${\bf \nu_j}$ correspond to the stoichiometry vector of $j$-th reaction such that $ \nu_{i,j}$
is the stoichiometric coefficient of species $S_i$ in reaction $R_j$. Given that the reaction
network at time $t$ is in state ${\bf X}(t) = {\bf x}$, the propensity function, $a_j({\bf x})$,
is defined so that the infinitesimal quantity $a_j({\bf x})dt$ gives the transition probability 
of the $j$-th reaction to occur in the time interval $[t,t+dt]$. Propensities are typically
dependent on the state of the system and the reaction conditions (i.e., external parameters)
of the network such as temperature, pressure, etc.
Mathematically, $\{{\bf X}(t)\}_{t\in\mathbb R_+}$ is a continuous-time, time-homogeneous,
jump Markov process with countable state space $E\subset \mathbb N^N$. The transition
rates of the Markov process are the propensity functions $a_j(\cdot),\ j=1,...,M$. The
transition rates determine the clock of the updates
(or jumps) from a current state ${\bf x}$ to a new (random) state ${\bf x}'$%={\bf x} + {\bf \nu_{j}}$
through the total rate $a_0({\bf x}):=\sum_{j=1}^M a_j({\bf x})$
% which is the intensity
%of the exponential waiting (or sojourn) time for a jump from state ${\bf x}$
while the transition probabilities of the process are determined by the ratio $\frac{a_j({\bf x})}{a_0({\bf x})}$.
We refer to {\em Algorithm~$1$} for the details of the stochastic simulation.
%There are exact algorithms for the simulation of the reaction network such as
%stochastic simulation algorithm (SSA) of Gillespie \cite{Gillespie:76, Gillespie:77}
%or next reaction algorithm of Gibson and Bruck \cite{Gibson:00} as well as
%approximation algorithms such as $\tau$-leap \cite{Gillespie:01} and several
%variations of it \cite{Rathinam:03, Chatterjee:05}. As a demonstration, given that the
%system is at the state ${\bf X}(t) = {\bf x}$ at time $t$, SSA computes the waiting
%time $\Delta t$ as a random number drawn from an exponential distribution with the total
%rate $a_0({\bf x})$ as parameter while the $R_{j^*}$ reaction occurs where $j^*\in\{1,...,M\}$
%is chosen such that $\sum_{j=1}^{j^*-1}p_j({\bf x}) < u < \sum_{j=j^*}^{M}p_j({\bf x})$
%where $u$ is a random number uniformly chosen in the interval $[0,1]$. The
%new state is given by ${\bf X}(t+\Delta t) = {\bf x}' = {\bf x} + v_{j^*}$.

\subsubsection*{Relative Entropy}
Assume that two probability distributions (or more generally  probability measures) $\mathcal P$
and $\tilde{\mathcal P}$ have corresponding probability densities $p=p(x)$ and $\tilde p=\tilde p(x)$.
Then, the Relative Entropy or Kullback-Leibler divergence of $\mathcal P$ with respect to
$\tilde{\mathcal P}$ is defined as \cite{Kullback:59, Cover:91}
\begin{equation}\label{rel_entropy}
\RELENT{\mathcal P}{\tilde{\mathcal P}} :=
\int p(x)\log\left(\frac{p(x)}{\tilde p(x)}\right)dx \ .
\end{equation}
In a more general setting, relative entropy is defined as $\RELENT{\mathcal P}{\tilde{\mathcal P}} :=
\int \log\left(\frac{d \mathcal P}{d\tilde{\mathcal P}}\right)\, d \mathcal P$ where
$\frac{d \mathcal P}{d\tilde{\mathcal P}}$ is a function known as Radon-Nikodym derivative
while the integration is performed with respect to the probability measure $\mathcal P$, \cite{Kipnis:99}. 
A necessary
condition for the relative entropy to be well-defined is that the Radon-Nikodym derivative
exists which is satisfied when $\mathcal P$ is absolutely continuous with respect to
$\tilde{\mathcal P}$.
Relative entropy has been utilized in a diverse range of scientific fields from statistical
mechanics \cite{Kipnis:99} to coding in telecommunications (information theory) \cite{Cover:91}
and finance \cite{Avellaneda:97}, and it possesses the following three fundamental  properties:
\begin{itemize}
\item[(i)] it is always non-negative,
\item[(ii)] it equals to zero if and only if $\mathcal P= \tilde{\mathcal P}$ $\mathcal P$-almost
everywhere, and,
\item[(iii)] $\RELENT{\mathcal P}{\tilde{\mathcal P}}<\infty$ if and only if $\mathcal P$
and $\tilde{\mathcal P}$ are absolutely continuous with respect to each other.
\end{itemize}
From an information theory perspective, relative entropy quantifies the loss of information
when $\tilde{\mathcal P}$ is utilized instead of $\mathcal P$, \cite{Cover:91}.  In other words,
relative entropy quantifies the inefficiency of assuming an incorrect or perturbed distribution
$\tilde{\mathcal P}$ instead of employing  the true distribution $\mathcal P$. Therefore, even
though not a metric, relative entropy has been used as a suitable  quantity for the assessment
of parametric sensitivities since the higher the relative entropy (i.e., the information loss) in
some perturbed direction, the larger the sensitivity should be in this direction.

\subsubsection*{Pathwise Relative Entropy and Relative Entropy Rate}
Proceeding to  the pathwise formulation of the relative entropy, we
assume that the propensities depend on a parameter vector $\theta\in\mathbb R^K$
(i.e., $a_j({\bf x})\equiv a_j^{\theta}({\bf x})$) while the continuous-time jump
Markov process $\big\{{\bf X}(t)\big\}_{t\in\mathbb R_+}$
lies in the {\em stationary regime}. We denote by $\mu^{\theta}({\bf x})$ the steady
state (or stationary) distribution of the  stochastic process ${\bf X}(t)$.
%$t\in\mathbb R_+$ the distribution of the random variable  is the stationary (or invariant) distribution
%denoted by  which is independent of the time. 
The stationary path distribution of the process in the interval $[0,T]$ is denoted by $\PATHS$.
Notice that path distributions (i.e., time-series distributions) are high-dimensional complex
objects; for instance,  if we consider the simpler discrete-time Markov chain case
$\{{\bf Z}_n\}_{n\in\mathbb Z_+}$, defined by the transition probability density $p({\bf z},{\bf z}')$, then,
utilizing repeatedly the Markov property, the stationary path distribution of the time-series
$({\bf z}_0, {\bf z}_1, \dots, {\bf z}_T)$ is given by
\begin{equation*}\label{path-measure}
Q_{[0,T]}(\{{\bf Z}_n={\bf z}_n\}_{0\le n\le T})=\mbox{Prob}({\bf z_0,\dots,z_T}) = \mu({\bf z}_0) p({\bf z}_0,{\bf z}_1)\dots p({\bf z}_{T-1}, {\bf z}_T)\, .
\end{equation*}
Proceeding, we consider another continuous-time jump Markov process $\{\tilde{\bf X}(t)\}_{t\in\mathbb R_+}$
defined by perturbing the propensity functions by a small vector $\epsilon\in\mathbb R^K$.
The corresponding steady state  and path distributions of $\{\tilde{\bf X}(t)\}_{t\in\mathbb R_+}$
are denoted by $\mu^{\theta+\epsilon}({\bf x})$ and $\PATHSAPP$, respectively.
Let the two path distributions $\PATHS$ and $\PATHSAPP$ be absolutely continuous
with respect to each other which is satisfied when $a_j^{\theta}({\bf x})=0$ if and
only if $a_j^{\theta+\epsilon}({\bf x})=0$ holds for all ${\bf x}\in E$ and $j=1,...,M$.
Then, the Relative Entropy of the path distribution $\PATHS$ with respect to  $\PATHSAPP$
is defined similarly to \VIZ{rel_entropy} as
\begin{equation}
\label{relent0}
\RELENT{{\PATHS}}{{\PATHSAPP}} :=
\int \log\left(\frac{d{\PATHS}}{d{\PATHSAPP}}\right)\, d\PATHS \, ,
\end{equation}
where $\frac{d{\PATHS}}{d{\PATHSAPP}}$ is the Radon-Nikodym derivative of
$\PATHS$ with respect to $\PATHSAPP$. In fact, using  the Girsanov's formula, we can
obtain an explicit expression for the Radon-Nikodym derivative %$d\PATHS/d\PATHSAPP$
in terms of the propensities, \cite{Kipnis:99}.
In the context of sensitivity analysis, 
the  pathwise relative entropy $\RELENT{{\PATHS}}{{\PATHSAPP}}$  is a measure of  information loss
 due to an $\epsilon$-perturbation
of the model parameters,  and consequently it is a natural measure of parametric sensitivity.
%Another important property of the path-wise relative entropy $\RELENT{{\PATHS}}{{\PATHSAPP}}$
%is that it is a non-decreasing function of time. 

Moreover, in the stationary regime, relative entropy
increases linearly in  time, hence the  Relative Entropy Rate (RER)
which is the time average of the relative entropy,
\begin{equation}\label{RER:Th}
\RELENTR{{Q^{\theta}}}{{Q^{\theta+\epsilon}}} := \lim_{T\rightarrow\infty} \frac{1}{T} \RELENT{{\PATHS}}{{\PATHSAPP}}\ ,
\end{equation}
is a well-defined quantity, \cite{Dumitrescu:88}. As first proposed in \cite{Pantazis:Kats:13},
$\RELENTR{{Q^{\theta}}}{{Q^{\theta+\epsilon}}}$ is
a suitable time-independent measure of sensitivity:
it measures the rate  of the loss of information due to an $\epsilon$-perturbation
of the model parameters, in the long-time,  stationary  dynamics regime of the
stochastic process. Furthermore, RER admits an
explicit formula given by (see Supplementary Information for a rigorous derivation)
\begin{equation}\label{RER:MP}
\RELENTR{{Q^{\theta}}}{{Q^{\theta+\epsilon}}}
= \mathbb E_{\mu^\theta} \Big[\sum_{j=1}^M a_j^{\theta}({\bf x})
\log \frac{a_j^{\theta}({\bf x})}{a_j^{\theta+\epsilon}({\bf x})}
-(a_0^{\theta}({\bf x}) - a_0^{\theta+\epsilon}({\bf x})) \Big]\, .
\end{equation}
Thus, from a practical point of view, RER is  an observable of the stochastic process
which can be computed numerically as an ergodic average, requiring only the knowledge
of the propensity functions and the stoichiometric matrix $(\nu)_{i, j}$. Nevertheless, in
order to carry out  the sensitivity analysis in  the parameter vector $\theta$, the computation
of RER for different $\epsilon$'s is necessary which can  be computationally challenging
for high-dimensional parameter spaces. Thus, a sensitivity analysis methodology which
does not depend on $\epsilon$'s --such methods are called ``gradient-free"-- is desirable
and is developed next.

\subsubsection*{Pathwise Fisher Information Matrix}
Even though not  directly evident from \VIZ{RER:MP}, a Taylor series expansion of RER
 in terms of  $\epsilon$ reveals that RER is locally a quadratic
function of the parameter vector $\epsilon\in\mathbb R^K$. Indeed, RER is non-negative
when $\epsilon\neq0$ and equals to zero when $\epsilon=0$ thus the linear term in
the Taylor expansion is zero. Therefore, RER is written --under  smoothness assumptions
on the propensity functions in the parameter vector $\theta$-- as \cite{Pantazis:Kats:13}:
\begin{equation}
\RELENTR{{Q^{\theta}}}{{Q^{\theta+\epsilon}}} = \frac{1}{2} \epsilon^T \FISHERR(Q^{\theta}) \epsilon + O(|\epsilon|^3)\, ,
\label{GFIM}
\end{equation}
where $\FISHERR(Q^{\theta})$ is a  $K\times K$ matrix that can be considered
as a pathwise analogue for the steady state Fisher Information Matrix (FIM). Similarly 
to  the steady state FIM for parametrized distributions \cite{Cover:91}, $\FISHERR(Q^{\theta})$
is the Hessian of the RER which geometrically corresponds to the curvature around
the minimum value of the relative entropy rate.
The pathwise FIM contains up to third order accuracy all the sensitivity information for the path
distribution at point $\theta$ for any perturbation direction $\epsilon$, therefore, 
the computation of the FIM is sufficient up to third order for the evaluation
of all the local sensitivities of the path distribution around the parameter vector $\theta$.
Moreover, an explicit formula for the pathwise FIM is given by (see Supplementary Information for a derivation)
\begin{equation}
\label{FIM:MP}
\FISHERR(Q^{\theta}) := 
\mathbb E_{\mu^{\theta}}\left[ \sum_{j=1}^M a_j^\theta({\bf x})
\nabla_\theta \log a_j^\theta({\bf x}) \nabla_\theta \log a_j^\theta({\bf x})^T \right] \ .
\end{equation}
The implications of this explicit formula are twofold. First, it reveals that for many
typical reaction networks the FIM has
a special block-diagonal structure which reflects the parameter interdependencies
and it is discussed in detail below. Second,
the FIM is based on the propensity functions as well as on their derivatives which
are known --actually, they define the process-- thus the  FIM,  similarly to RER,
is numerically computable as an observable of the process. Subsequent sections
present various strategies  to numerically estimate both the RER and the FIM in
an efficient fashion .

Furthermore, the pathwise FIM, $\FISHERR(Q^{\theta})$, can be used  not only
for the estimation/approximation of RER via \VIZ{GFIM} but also to infer intrinsic
knowledge for  the system's sensitivities \cite{Emery:98, Majda:10}. In general, the
spectral analysis of the FIM  reveals the (local) most/least sensitive directions
of the system around $\theta$. Indeed, by ordering the eigenvalues of the FIM as
$$\lambda_1^\theta\geq...\geq\lambda_K^\theta\geq0\, ,$$
it can be inferred that the most sensitive direction corresponds to the eigenvector
with eigenvalue $\lambda_1^\theta$ while the least sensitive direction corresponds
to the eigenvector with eigenvalue $\lambda_K^\theta$. Additionally, the FIM is one
of the most useful tools for optimal experimental design. Many of the optimality
criteria such as D-optimality where the determinant of the FIM is maximized or
A-optimality where the trace of the inverse of the FIM is minimized
are based on FIM, \cite{Emery:98}. 
%Even
%though the proposed pathwise FIM is not derived from a statistical description
%of measured data, meaning that the measurement noise is not taken into account,
%but from the stochastic model itself, it can be utilized for the optimal design of
%experiments to account the intrinsic properties of the system. 
In the same direction,
robustness of the system to parameter perturbations or errors as well as parameter
identifiability can be studied utilizing spectral analysis of the FIM. For
instance, parameter identifiability is satisfied when all the eigenvalues of the FIM
are above a given threshold, \cite{Komorowski:11}.

\subsubsection*{Sensitivity analysis at the logarithmic scale}
\label{log:sens:analysis:app}
In many  biochemical reaction networks, the model parameters differ
by orders of magnitude and a reasonable option for carrying out sensitivity analysis
is to perform perturbations which are proportional to the parameter magnitude.
This can be carried out  by perturbing the logarithm
of the model parameters instead of the parameters itself. Using the 
chain rule  $\nabla_{\log\theta} f(\theta) = 
\nabla_{\theta} f(\theta) . \nabla_{\log\theta} \theta = \theta . \nabla_{\theta} f(\theta)$
where `$.$' is defined as the  element by element multiplication (i.e., $(a.b)_k=a_kb_k, \ k=1,...,K$), we obtain
the logarithmically-scaled FIM:
\begin{equation}\label{propensities:general}
\big( \FISHERR({Q^{\log\theta}}) \big)_{k,l} = \theta_k \theta_l \big( \FISHERR({Q^{\theta}}) \big)_{k,l}
\ ,\ \ \ \ k,l=1,...,K \ ,
\end{equation}
where  $\FISHERR({Q^{\theta}})$ is given by \VIZ{FIM:MP}.
Similarly, the logarithmic perturbation for the RER is carried out  using  the
perturbation vector $\theta . \epsilon$ instead of $\epsilon$.
Notice that \VIZ{GFIM} continues to be valid for the logarithmic scale, i.e.,
\begin{equation}
\RELENTR{{Q^{\theta}}}{{Q^{\theta+\theta.\epsilon}}} = \frac{1}{2}\epsilon^T
\FISHERR({Q^{\log\theta}}) \epsilon + O(|\epsilon|^3) \ .
\end{equation}

%%%%%%%%%%%%%%%%%%%%%%%%%%%%%%%%%%%%%%%%
\subsubsection*{Linking relative entropy and observables}
As we discussed in the previous sections,  relative entropy provides a mathematically elegant
and computationally tractable  methodology for the parameter sensitivity analysis of complex,
stochastic dynamical systems. Such  results focus on the sensitivity of the entire probability
distribution, either at equilibrium or at the path-space level, i.e., for the entire stationary time-series.
However, in most simulations of chemical and biological networks, the main focus of interest is
observables such as mean populations, population correlations, population variance as well
as path-space observables such as time autocorrelations and extinction times. Therefore, it is
plausible to attempt to connect the parameter sensitivities of  observables to the relative entropy
methods proposed here. Indeed, relative entropy can provide an upper bound for a large family
of observable functions through the Pinsker (or Csiszar-Kullback-Pinsker) inequality, \cite{Cover:91}. 
%The Pinsker inequality
%states that the total variation norm between probability distributions $Q^{\theta}$ and $Q^{\theta+\epsilon}$
%is bounded from above by the square root of twice the relative entropy \cite{Cover:91}.
More precisely, for any  bounded observable function $f$, the Pinsker inequality states that
\begin{equation}
|\mathbb E_{Q^{\theta}}[f] - \mathbb E_{Q^{\theta+\epsilon}}[f]| \leq ||f||_\infty \sqrt{2 \RELENT{Q^{\theta}}{Q^{\theta+\epsilon}}}\, ,
\label{Pinsker:ineq}
\end{equation}
where $||\cdot||_\infty$ denotes  the supremum (here, maximum) of $f$.
%under the assumption that both $\mathcal P$ and $\tilde{\mathcal P}$ have densities. 
An obvious outcome of this  inequality is that if the (pseudo-)distance between two 
distributions defined  by $\RELENT{Q^{\theta}}{Q^{\theta+\epsilon}}$ is controlled,
then the error between the two distributions is also controlled for any bounded
observable.
In the context of sensitivity analysis, inequality \VIZ{Pinsker:ineq} states that if the
relative entropy is small, i.e., insensitive in a particular parameter  direction, then,
any  bounded observable $f$ is also expected to be  insensitive towards the same
direction. In this sense, \VIZ{Pinsker:ineq} can be viewed as a ``conservative'' --but
not necessarily  sharp-- bound for the parametric sensitivity analysis of observables, including 
path-dependent observables such as  long-time averages and autocorrelations.

From a practical perspective, \VIZ{Pinsker:ineq} can be used as an indicator that suggests
--even in the presence of a very high-dimensional parameter space-- which are the  insensitive
parameter directions for observables of stochastic dynamical systems, based on the proposed
relative entropy methods. The suggested least-sensitive  parameter-space directions can be
subsequently  verified computationally. We provide two examples of  this practical strategy
in the p53 and the EGFR models discussed in the sequel.

\medskip
\noindent
{\bf Remark 1:} 
However, we note  that  in order to carry out such an  analysis  in a mathematically rigorous
manner, we need to require that the norm  $||\cdot||_\infty$ in \VIZ{Pinsker:ineq}  is controlled.
For instance, typical observables in biochemical reaction networks are  the number of molecules
for each  species, hence  $f(x)=x$. Thus, for reaction networks where the population size is
large   the Pinsker inequality \VIZ{Pinsker:ineq} will provide a  bound that may  not  be sharp. 
Nevertheless, it is possible to derive a tighter inequality where the maximum norm of $f$ in
\VIZ{Pinsker:ineq} is replaced by the standard deviation of $f$, we refer to the forthcoming
publication \cite{Kats:Pantazis:prep} and to recent related work on uncertainty quantification
and relative entropy, \cite{Dupuis:13}.

\subsection*{Block-diagonal structure of the pathwise FIM}
\label{FIM:structure:sec}
In chemical reaction networks,  reactions typically  depend only on
a small subset of the parameter vector. Mathematically, this is described as
\begin{equation}
a_j^\theta({\bf x})=a_j({\bf x};\theta_{k_1},...\theta_{k_{L_j}})\, ,
\label{propensities:sparse}
\end{equation}
where $k_1,...,k_{L_j}\in\{1,...,K\}$ while $L_j\ll K$ is the number of involved parameters in reaction $R_j$.
Using \VIZ{FIM:MP},  it can be shown that this parametric dependence of the propensities
is directly reflected on the  FIM. Indeed, after grouping the reactions into subsets
in such a way that each subset contains the minimum number of reactions having
common parameters, the pathwise FIM --upon rearrangement of the parameter vector--
becomes a block-diagonal matrix. The pathwise FIM is then written as
\begin{equation}\label{Block-diagonal}
\FISHERR({Q^{\log\theta}}) = \left[\begin{array}{ccc} A_1^\theta & & \text{\Large0} \\
& \ddots &  \\ \text{\Large0} & &  A_I^\theta \end{array} \right]
\end{equation}
where $A_1^\theta,...,A_I^\theta$ are block matrices. The block matrices
which are defined by the reaction subsets with the same parametric
dependence are easily obtained by creating a graph whose nodes are the
reactions and the parameters while the edges are their dependences. Then,
the parameter nodes contained in a connected subgraph define a parameter
subset which in turn corresponds to a block of the FIM. An illustration of
this procedure is shown in Figure~\ref{FIM:structure} where a reaction
network with $M=9$ reactions and $K=7$ parameters is plotted. The parametric
dependencies of the reactions are shown in the left panel where 4 subgroups
of parameters are defined based on the graph connectivity. The resulting
block-diagonal structure of the FIM is shown on the right panel  of Figure~\ref{FIM:structure}.

\begin{figure}[!htb]
\begin{center}
\includegraphics[width=.6\textwidth]{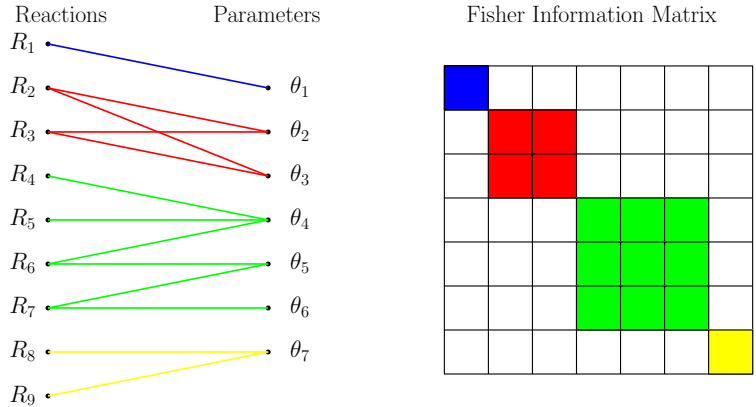}
\caption{Left panel: The graph representation of the dependencies between the reactions
(left column) and the model parameters (right column). The grouping of the parameters is then
carried out  by noting  the connected parts of the graph. Right panel: The corresponding
block-diagonal structure of the FIM. In this example $K=7$ while the largest dimension of the
blocks is $L=3$.}
\label{FIM:structure}
\end{center}
\end{figure}

Before proceeding with the theoretical computation of the FIM for various well-known
classes of biochemical reaction networks, we list some of the implications
of this simplified structure of the FIM in sensitivity analysis and elsewhere.
\begin{itemize}
\item[(i)] The sparsity of the FIM is proportional to the parametric decoupling between the
reactions. Knowing a priori the zero elements of the FIM, there is no need to numerically
compute them. It is clear that the computation cost for each sample drops from $O(K^2)$
to $O(KL)$ where $L$ is the largest dimension of the block matrices.

\item[(ii)]  The inverse of the FIM is also
block-diagonal and each block of the inverse FIM is the inverse of the respective
block. This fact allows us in the parameter estimation problem  to easily evaluate   the  lower bound of the
variance, at least  for the complete-data case \cite{Wilkinson:12}, i.e., obtain Cramer-Rao bounds,
\cite{Kay:93, Wasserman:04} which are given by the diagonal elements of the
inverse of the FIM.

\item[(iii)] Relation \VIZ{Block-diagonal} implies that that optimality criteria in  optimal
experiment design, \cite{Emery:98}, are significantly simplified.  For example, the determinant
employed  in the D-optimality test is given by the relation  $\det (\FISHERR) = \prod_{i=1}^I \det(A_i)$,  while the trace of the
inverse of the FIM utilized in the A-optimality reduces to  $\text{tr} (\FISHERR^{-1}) = \sum_{i=1}^I \text{tr}(A_i^{-1})$.
%Moreover, the designer of an experiment may
%try to handle each subset of parameters differently and if possible independently by
%designing  specialized sub-experiments associated  to specific  parameter subsets. Notice, however,
%that the parameter subsets are not fully independent because of the dependencies with
%the species which should be taken into account, too. 

\item[(iv)] Given that parametric identifiability is characterized by the magnitude of the
eigenvalues of the FIM, e.g. a zero eigenvalue corresponds to a non-identifiable direction
in parameter space, \cite{Rothenberg:71, Komorowski:11}, then the block-diagonal structure
\VIZ{Block-diagonal}  can provide additional  insights in parameter identification.
For instance, the identifiability of the parameters of the group that corresponds to the $i$-th
block, $A_i$, will be increased if  the smaller eigenvalues of the $i$-th block can be increased.
%, e.g. note the differences in identifiability between  stationary and transient FIM in Figure~\ref{EGFR:FIM:fig}.
On the other hand, if the determinant of the $i$-th block equals to zero then at least one of
its eigenvalues is zero and thus the corresponding linear combinations of  parameters are
non-identifiable.  Similarly, the robustness of the system to perturbations of the parameters
of the $i$-th group will be increased if there is a way to decrease the larger eigenvalues of
the $i$-th block.

% For instance, the identifiability of the
%$k$-th parameter in the subsequent example will be increased if there is a way to
%increase the $k$-th diagonal element of the FIM. Analogously, the robustness
%of the system to the $k$-th parameter in the same example will be increased if the
%$k$-th diagonal element of the FIM is  decreased accordingly.
\end{itemize}
Overall,  we note that extracting   useful information regarding   model parameters
can be performed for each block of the pathwise FIM independently.
Next, we discuss two specific examples of biochemical reaction networks where the
explicit calculation of the block-diagonal FIM is demonstrated.

\subsubsection*{Reactions with independent reaction constants}
\label{spe:case}
An important class of well-mixed reaction networks take  the general form ``$\alpha_j A_j+\beta_j B_j\overset{\theta_j}{\rightarrow}...$''
where $A_j$ and $B_j$ are the reactant species while $\alpha_j$ and $\beta_j$ are the
respective number of molecules needed for the reaction.
The reaction constant, $\theta_j$, is the parameter of the $j$-th reaction.
The propensity function for the $j$-th reaction is given as the product between
a rate constant and a function of the current state ${\bf x}$:
\begin{equation}
a_j({\bf x}) = \theta_j g_j({\bf x}),\ \ \ j=1,...,M \ .
\label{special:propensities}
\end{equation}
Typically, $g_j({\bf x}) = \binom{{\bf x}_{A_j}}{\alpha_j} \binom{{\bf x}_{B_j}}{\beta_j}$ which
stems from the law of mass action, however,
it can take different forms depending on the modeling of the physical process. This
reaction network has $K=M$ parameters, while each propensity depends only on
one parameter, i.e., $L_j=1$ in \VIZ{propensities:sparse} for $j=1,...,M$.
The $(k,l)$-th   element of the FIM in the logarithmic scale is explicitly given by
\begin{equation}\label{FIM:log}
\big( \FISHERR({Q^{\log\theta}}) \big)_{k,l} = \theta_k \theta_l \mathbb E_{\mu^{\theta}}
\left[ \sum_{j=1}^M a_j^\theta({\bf x}) \partial_{\theta_k} \log a_j^\theta({\bf x}) \partial_{\theta_l} \log a_j^\theta({\bf x})^T \right] \ ,
\end{equation}
where $\mu^{\theta}$ is the stationary distribution of the stochastic process.
Furthermore, it holds %obtained for propensity functions of type \VIZ{special:propensities} 
that
$\partial_{\theta_k} \log a_j^\theta({\bf x}) = \frac{1}{\theta_k} \delta_k(j)$ where $\delta(\cdot)$
is the Dirac function, therefore the pathwise FIM is a diagonal matrix with elements given by
\begin{equation}
\big( \FISHERR({Q^{\log\theta}}) \big)_{k,l} =  \left\{\begin{array}{cc}
\mathbb E_{\mu^{\theta}}\big[ a_k^\theta({\bf x}) \big] \ , & l=k \\
0 \ , & l\neq k \end{array}\, . \right.
\label{diagonal}
\end{equation}
This result  demonstrates  that the sensitivity of a reaction constant is proportional to
the equilibrium average of the respective propensity function. Moreover, due to
the diagonal form of the FIM, it is straightforward to carry out the  eigenvalue analysis
and infer  the most/least sensitive directions of the reaction network:
the eigenvalues of the FIM are its diagonal elements while the eigenvectors
are the standard basis vectors of $\mathbb R^K$. Hence, the most (respectively least) sensitive
parameter is obtained from the largest (respectively  smallest) diagonal element of the FIM.
Furthermore, \VIZ{diagonal} demonstrates that the (local) robustness of the reaction network to a
specific parameter is inversely proportional to  the mean  propensity of the corresponding reaction.
Another observation stemming from the diagonal structure of the pathwise FIM is that each
rate constant can be estimated from time-series data independently from the other rate
constants. This observation has been already pointed out and discussed in the context of
maximum likelihood estimation for the complete-data case \cite[Sec. 10.2]{Wilkinson:12}.

Additionally, the mean  firing rate of a reaction is equal  to the mean  propensity.
Hence, it can be stated that the parameters that correspond to the faster reactions, i.e., to reactions
with larger mean  firing rate, are more sensitive in a pathwise entropy sense. It should be noted,
however, that not all observables are sensitive to the parameters that correspond to the faster reactions
and there are examples (see the protein production-degradation model in the Results section) where
steady state observables such as the equilibrium distribution  remain insensitive to specific perturbation directions even though their
mean  propensity may be increased, see \VIZ{equilibrium_FIM} and \VIZ{diagonal:example1}.

 Finally, we would like to remark that even though
$\mathbb E_{\mu^{\theta}}\big[ a_k^\theta({\bf x}) \big] = \theta_k\mathbb E_{\mu^{\theta}}\big[ g_k^\theta({\bf x}) \big]$ trivially holds true,
the diagonal elements of the FIM are not linear functions of the corresponding
reaction constants since  the steady state  distribution $\mu^\theta$, depends also
on the parameter vector $\theta$. In fact, high reaction constants
do not necessarily imply large mean  propensities and hence a more sensitive parametric
dependence. This is  specifically due to the mean value in \VIZ{diagonal} and 
as an illustrative example we refer to the simple protein production-degradation model
(e.g., compare \VIZ{Ex1:propensities} and \VIZ{diagonal:example1}).

\subsubsection*{Michaelis-Menten kinetics}
Another important class of reaction networks is the Michaelis-Menten kinetics.
In its simplest form, this chemical network contains two reactions between species $A$ and $B$
(i.e., $A\leftrightarrow B$) with propensity functions given by
\begin{equation*}
a_k^\theta({\bf x}) = \frac{\theta_k{\bf x}_A}{\theta_{k+1}+{\bf x}_A}\ \ \ \ \ \ \ \text{and} \ \ \ \ \ \ \ 
a_{k+1}^\theta({\bf x}) = \frac{\theta_k{\bf x}_B}{\theta_{k+1}+{\bf x}_B} \ .
\end{equation*}
This reaction sub-network which is derived under a quasi-steady-state assumption is one of
the best-known models of enzyme kinetics in biochemistry. The parameter
$\theta_k$ (usually denoted by $V_{\max}$) represents the maximum rate achieved by the
system, at maximum (saturating) substrate concentrations while the Michaelis constant $\theta_{k+1}$
(usually denoted by $K_m$) is the substrate concentration at which the reaction rate is half the
maximum value.
The propensities of this Michaelis-Menten sub-network depend on two parameters
($L_k=L_{k+1}=2$ in \VIZ{propensities:sparse}) thus the corresponding FIM block is
a $2\times2$ matrix. The elements of the FIM matrix 
are given by
\begin{equation}
\big( \FISHERR({Q^{\log\theta}}) \big)_{k,l} =  \left\{\begin{array}{cc}
\mathbb E_{\mu^{\theta}}\big[ a_k^\theta({\bf x}) + a_{k+1}^\theta({\bf x}) \big] \ , & l=k \\
-  \mathbb E_{\mu^{\theta}}\big[ a_k^\theta({\bf x})\frac{\theta_{k+1}}{\theta_{k+1}+{\bf x}_A}
+ a_{k+1}^\theta({\bf x}) \frac{\theta_{k+1}}{\theta_{k+1}+{\bf x}_B} \big] \ , & l=k+1 \\
0 \ , & l\neq k,k+1 \end{array} \right.
\end{equation}
for the $k$-th row while the $k+1$-th row is given by
\begin{equation}
\big( \FISHERR({Q^{\log\theta}}) \big)_{k+1,l} =  \left\{\begin{array}{cc}
\mathbb E_{\mu^{\theta}}\big[ a_k^\theta({\bf x}) \frac{\theta_{k+1}^2}{(\theta_{k+1}+{\bf x}_A)^2}
+ a_{k+1}^\theta({\bf x}) \frac{\theta_{k+1}^2}{(\theta_{k+1}+{\bf x}_B)^2} \big] \ , & l=k+1 \\
-  \mathbb E_{\mu^{\theta}}\big[ a_k^\theta({\bf x})\frac{\theta_{k+1}}{\theta_{k+1}+{\bf x}_A}
+ a_{k+1}^\theta({\bf x}) \frac{\theta_{k+1}}{\theta_{k+1}+{\bf x}_B} \big] \ , & l=k \\
0 \ , & l\neq k+1,k  \ \ . \end{array} \right.
\end{equation}
In general,  biochemical reaction networks may have significantly more complex propensities,
nevertheless, the computation of the FIM elements follows exactly the same calculation lines
for any propensity function.

\subsection*{Strategies for the statistical estimation of RER and FIM}
\label{stat:estim:sec}
Previous sections introduced and justified RER and FIM as appropriate observables for
measuring the sensitivity analysis of the reaction network's parameters in long-time dynamics.
This section presents  strategies on how to efficiently estimate these quantities as
ergodic averages of the underlying stochastic process.

\subsubsection*{Unbiased Statistical estimators}
Since the stationary distribution, $\mu^\theta$, is usually not known,
both FIM and RER should be estimated numerically as ergodic averages. Indeed, the statistical
ergodic estimator for RER is given by
\begin{equation}
\bar{\mathcal H}^{(n)} =  \frac{1}{T} \sum_{i=0}^{n-1} \Delta t_i \Big[ \sum_{j=1}^M a_j^\theta({\bf x}_i)
\log \frac{a_j^\theta({\bf x}_i)}{a_j^{\theta+\epsilon}({\bf x}_i)}
- \big(a_0^\theta({\bf x}_i) - a_0^{\theta+\epsilon}({\bf x}_i)\big) \Big]
\label{RER:num:approx:MP}
\end{equation}
where $\Delta t_i$ is an exponential random variable with parameter given by the total rate, $a_0^\theta({\bf x}_i)$,
while $T=\sum_{i=1}^n \Delta t_i$ is the total simulation time. The sequence $\{{\bf x}_i\}_{i=0}^n$
is the embedded Markov chain with transition probabilities from state ${\bf x}_i$ to state ${\bf x}_{i+1}$
is given by the ratio $\frac{a_j^\theta({\bf x}_i)}{a_0^\theta({\bf x}_i)}$. The weight $\Delta t_i$, which is the waiting time at state ${\bf x}_i$, 
 is necessary for the
unbiased estimation of the observable, \cite{Gillespie:76}. Similarly, the unbiased estimator for the
FIM is 
\begin{equation}
\bar{\bf F}_{\mathcal H}^{(n)} = \frac{1}{T} \sum_{i=0}^{n-1} \Delta t_i \sum_{j=1}^M 
a_j^\theta({\bf x}_i) \nabla_\theta \log a_j^\theta({\bf x}_i) \nabla_\theta \log a_j^\theta({\bf x}_i)^T \ .
\label{FIM:num:approx:MP}
\end{equation}
Noticing that the computation of the local propensity functions $a_j^\theta({\bf x}_i)$ for
all $j=1,...,M$ is needed for the simulation of the jump Markov process when Monte Carlo
methods such as SSA \cite{Gillespie:76} is utilized,
the computation of the perturbed transition rates is the only additional computational
cost for the numerical RER while the additional cost for the estimation of the FIM is
the computation of the derivatives of the propensities. {\em Algorithm 1} summarizes
the numerical computation of RER and FIM, employing the  SSA for the simulation
of the jump Markov process.

\vspace{5mm}
{\em Algorithm 1: SSA-based numerical computation of RER and FIM.}
\begin{enumerate}
\item Initialize: ${\bf x}={\bf x}_0$, $T=0$, $\bar{\mathcal H}=0$ and $\bar{\bf F}=0$.
\item FOR $i=1,...,n$
\begin{enumerate}
\item Compute: $\big\{a_j^\theta({\bf x})\big\}_{j=1}^M$, $a_0^\theta({\bf x})$.
Compute also $\big\{a_j^{\theta+\epsilon}({\bf x})\big\}_{j=1}^M$ (only for RER) and
$\big\{\nabla_\theta \log a_j^\theta({\bf x})\}_{j=1}^M$ (only for FIM).
\item $\Delta t = -\log(u_1)/a_0^\theta({\bf x})$ where $u_1\sim \mathcal U(0,1)$.
\item Update time: $T = T + \Delta t$
\item Update RER: $\bar{\mathcal H}=\bar{\mathcal H} + \Delta t \Big[ \sum_{j=1}^M a_j^\theta({\bf x})
\log \frac{a_j^\theta({\bf x})}{a_j^{\theta+\epsilon}({\bf x})}
- \big(a_0^\theta({\bf x}) - a_0^{\theta+\epsilon}({\bf x})\big) \Big]$.
\item Update FIM: $\bar{\bf F}=\bar{\bf F} + \Delta t \sum_{j=1}^M 
a_j^\theta({\bf x}) \nabla_\theta \log a_j^\theta({\bf x}) \nabla_\theta \log a_j^\theta({\bf x})^T$.
\item Find $j^*$ such that $\sum_{j=1}^{j^*-1}a_j({\bf x}) < u_2 a_0({\bf x}) < \sum_{j=j^*}^{M}a_j({\bf x})$ where $u_2\sim \mathcal U(0,1)$.
\item Update state: ${\bf x}={\bf x} + {\bf \nu_{j^*}}$.
\end{enumerate}
\item Normalize: $\bar{\mathcal H}=\bar{\mathcal H}/T$ and $\bar{\bf F}=\bar{\bf F}/T$.
\end{enumerate}

\subsubsection*{Accelerated statistical estimators}
\label{ode:approx}
A typical feature of biochemical systems is that the modeled reaction network
is large with hundreds or thousands of reactions and different time scales  stemming
from the orders of magnitude difference between the reaction rates and/or between the
species concentrations, making the SSA extremely slow.  A large number of
multi-scale approximations of the original SSA have been
developed in order to handle such issues resulting to accelerated simulation algorithms.
For example, mean-field approximation ignores the fluctuations of the stochastic process and yields a
deterministic system of ordinary differential equations (ODE)  for the mean population  of the species \cite{Gardiner:85,
vanKampen:06}. Stochastic corrections to the mean-field model such as stochastic
Langevin \cite{Gillespie:00} and linear noise approximation \cite{Kurtz:72} can be applied in order
to improve the accuracy of the simulation. An alternative approximation of the jump
Markov process is the tau-leap method proposed by Gillespie \cite{Gillespie:01} where
a batch of events occurs at each time-increment, $\tau$. Several improvements of
the basic tau-leap algorithm on how to select adaptively the $\tau$ \cite{Cao:06}
or  avoiding negative populations \cite{Tian:04, Chatterjee:05} have been proposed,
however, their performance is heavily model-dependent.

In this subsection, we propose such  approximations in order to
efficiently compute the FIM and/or RER observables, while maintaining controlled bias
in the statistical estimators. As an illustration, we present the well-known mean-field approximation. 
The popularity of the mean-field modeling stems from
their computational efficiency. To proceed, the stochastic process can be written as
\begin{equation}\label{approx:MFE}
{\bf X}(t) = x(t) + \eta \xi(t)
\end{equation}
where $x(t)$ is the deterministic part (mean) of the process, $\xi(t)$ is the stochastic zero-mean
part while $\eta$ is the amplitude of the stochastic term. The amplitude of the stochastic term
is proportional to the inverse square root of the reactant
populations \cite{Kurtz:72, Kurtz:81, Gillespie:00}. Thus, for large populations, the fluctuations
of the time-evolving species populations become vanishingly small compared to the deterministic
contributions. Consequently, the dominant part of the process is the deterministic term
whose dynamics are governed by the ODE system
\begin{equation}
\dot{x}_i(t) = \sum_{j=1}^M \nu_{j,i} a_j^\theta(x(t))\ ,\ \ i=1,...,N \ .
\label{mean:field:eq}
\end{equation}
This ODE system is also known as reaction rate equations \cite{Gillespie:00}.
Restricted for simplicity to the special case with independent rate constants
for each reaction, the diagonal elements of the FIM are approximated using
\VIZ{approx:MFE} as
\begin{equation}
\begin{aligned}
\big( \FISHERR({Q^{\log\theta}}) \big)_{k,k} &= \mathbb E_{\mu^{\theta}}\big[ a_k^\theta({\bf x}) \big]
\approx \frac{1}{T} \sum_{i=1}^{n} \Delta t_i a_k^\theta({\bf X}(t_i)) \\
&= \frac{1}{T} \sum_{i=1}^{n} \Delta t_i\, a_k^\theta\big(x(t_i) + \eta \xi(t_i)\big) \\
&= \frac{1}{T} \sum_{i=1}^{n} \Delta t_i\, a_k^\theta(x(t_i)) + O(\eta) \ 
\end{aligned}
\end{equation}
Typically, such ODE system is solved  using  an adaptive time-step numerical integrator
up to final time $T=\sum_{i=0}^n \Delta t_i$. Thus, for large species populations
($|S_i|\gg1$), the following numerical estimator for the FIM's diagonal elements is obtained:
\begin{equation}
\big( \bar{\bar{{\bf F}}}_{\mathcal H}^{(n)} \big)_{k,k} = \frac{1}{T} \sum_{i=1}^{n} \Delta t_i\, a_k^\theta(x(t_i)) \ \ , \ k=1,...,K
\label{spe:case:num:FIM}
\end{equation}
Relation \VIZ{spe:case:num:FIM} suggests an algorithm  similar to {\em Algorithm 1} for
the numerical computation of the FIM  where instead of SSA, an ODE solver is employed.

\medskip
\noindent
{\bf Remark 2:} Multi-scale approximations are usually
valid for large populations and relatively simple systems which do not exhibit complex dynamics
such as bistability or intermittency. Indeed, large deviation arguments \cite{Doering:07}
or even explicitly available formulas for escape times \cite{Hanggi:84} demonstrate that stochastic
approximations cannot always capture correctly exit times, rare events, strong intermittency, etc.
even in relatively simple systems.
However, in order to simulate large biochemical systems there is often no other alternative than
such approximate models, which nevertheless need to be employed with the necessary caution.

\medskip
\noindent
{\bf Remark 3:}  In biochemical systems, we  are interested not only in the steady state,
i.e., the stationary distribution  or time-series, but also in the transient regime, e.g. 
signaling phenomena. The extension of the proposed sensitivity
analysis method to the transient regime is justified by the fact that the time-normalized
relative entropy can be also decomposed as a sum of simple integrals \cite{Cover:91}
which results to the fact that the statistical estimators \VIZ{RER:num:approx:MP}
and \VIZ{FIM:num:approx:MP} remain  valid. In a subsequent section
we present an example of a biochemical system (EGFR) which exhibits transient behavior,
and where the proposed sensitivity analysis tools are tested and validated. The rigorous
mathematical derivation of the relative entropy rate for the transient regime is out of the scope of
this publication and a dedicated mathematical article on the time dependent relative entropy rate
will follow.

\section*{Results}
\label{results:sec}

\subsection*{A simple  protein production/degradation model }
\label{birth:death:sec}
We first consider an elementary stochastic model for protein production and degradation, \cite{Wilkinson:09},
which is also a component  of more complex models for gene regulatory networks, \cite{Thattai:01}.
In this simplified model, the protein is produced at a
constant rate $k_1$, while it is  degraded with rate $k_2$, corresponding 
to the  reactions 
\begin{equation}
\emptyset \underset{k_2}{\overset{k_1}{\rightleftarrows}} X \ .
\end{equation}
Accordingly, the corresponding propensity functions for the  current state ${\bf x}=x$ are:
\begin{equation}
a_1(x) = k_1\quad \text{and} \quad a_2(x)=k_2x \ .
\label{Ex1:propensities}
\end{equation}
We consider this simple stochastic model due to the available analytic representations of
the steady state (equilibrium) distribution, time-dependent moments and autocorrelations,
\cite[Sec. 7.1]{Gardiner:85}. Consequently, we can both  illustrate  
 the proposed pathwise sensitivity analysis, as well as  compare it 
to the  standard equilibrium FIM,  revealing concretely   differences
between the two approaches. 

The equilibrium distribution, $\mu^\theta$, of this simple network is a Poisson distribution with parameter
$\frac{k_1}{k_2}$. Therefore, the equilibrium FIM for the parameter
vector $\theta=[k_1,k_2]^T$ is given in logarithmic scale by
\begin{equation}\label{equilibrium_FIM}
\FISHER({\mu^{\log\theta}}) = \frac{k_1}{k_2} \left[\begin{array}{rr} 1 & -1 \\ -1 & 1 \end{array} \right] \, .
\end{equation}
On the other hand, the  pathwise FIM is computed via   \VIZ{diagonal}:
\begin{equation}\label{path_FIM}
\FISHERR({Q^{\log\theta}}) = k_1 \left[\begin{array}{rr} 1 & 0 \\ 0 & 1 \end{array} \right] \ , 
\end{equation}
where we used that 
\begin{equation}\label{diagonal:example1}
\mathbb E_{\mu^{\theta}}[a_1(x)]=E_{\mu^{\theta}}[a_2(x)]=k_1\, .
\end{equation}
The complete calculations can be found in the supplementary material.
Some of the implications of the differences between these two FIMs are discussed next.
%%%%%%%%%%%%%%%%%%%%%

First, we observe that  the equilibrium FIM, \VIZ{equilibrium_FIM},  is singular, i.e., one of the
eigenvalues is zero. We readily see that in the parameter direction  defined by the corresponding
eigenvector, i.e., when the parameter ratio, $\frac{k_1}{k_2}$, remains constant,
the system  is expected  to be insensitive, at least with respect to the equilibrium distribution.
Clearly, this is a fact verified directly from the Poisson equilibrium distribution $\mu^\theta$
which depends only on the ratio.
On the other hand, the pathwise FIM, \VIZ{path_FIM}, is not singular and all the directions are
equally sensitive. This fact  suggests
that  observables for  dynamic quantities  may  be   sensitive not only to parameter ratio
perturbations but also to other parameter perturbations. Indeed, one such example is
the stationary autocorrelation function, which in the case of the simple protein production/degradation
model is explicitly given by \cite[Sec. 7.1]{Gardiner:85},
\begin{equation}
\big<X_t,X_0\big>_{s} = \frac{k_1}{k_2} e^{-k_2 t}\, ,
\end{equation}
where $\big<\cdot,\cdot\big>_{s}$ denotes stationary averaging. Based on this formula,
it is obvious that  the autocorrelation
function is also sensitive to $k_2$, in addiction to the ratio $\frac{k_1}{k_2}$.
This example  demonstrates that in contrast to the pathwise FIM,
 the equilibrium FIM is inadequate to fully 
capture the dynamic properties of the process. Moreover, the pathwise FIM
depends linearly only on $k_1$, %and in the same way for both parameters, 
which shows  that
%: (a) both parameters as well as any other
%parameter direction are equally sensitive, at least regarding relative entropy,
%and (b) 
the reaction rate constants and propensity
functions in \VIZ{Ex1:propensities} alone, %or more generally  the terms $a_k^\theta({\bf x})$ in \VIZ{diagonal}, 
can be misleading in  the assessment of parametric sensitivity. Contrary, 
the mathematically correct  equilibrium averaging of the propensities, i.e., \VIZ{diagonal}
can lead to a completely different outcome, as can be readily seen when we 
compare  \VIZ{Ex1:propensities} and \VIZ{diagonal:example1}.

In terms of parameter identifiability, the fact that one of the eigenvalues of \VIZ{equilibrium_FIM}
is zero implies that  that the two-dimensional parameter vector of the system is
non-identifiable.  Indeed, the asymptotic normality of the maximum likelihood estimators,
\cite{Kay:93, Wasserman:04}, states that their variance  (also  a lower bound according to the
Cramer Rao theorem), which  determines parameter identifiability of $k_1$ and $k_2$,
is the reciprocal of the eigenvalues of \VIZ{equilibrium_FIM}. A straightforward calculation
involving the eigenvectors of \VIZ{equilibrium_FIM} shows that the only identifiable parameter
is the ratio of the reaction constants appearing in \VIZ{equilibrium_FIM}. Therefore parametric
inference for both parameters from equilibrium data is not possible. 
On the other hand, the pathwise FIM \VIZ{path_FIM}  is not singular, which readily implies
that both parameters can be identified through (complete) time-series data, provided that $k_1\neq0$.
Summarizing, this birth/death model is an example where equilibrium sampling is not enough
for the identifiability of all the parameters, however, if dynamics data are  available and are  taken
into account then all the parameters become identifiable as pathwise FIM asserts.
% and suggests that
%experimentalist should try to devise  new experiments where 
%when dynamics  information
%can be  measured,  and used for the improvement of the parameters' identifiability.
%%
%Lastly, the  pathwise FIM can be  further   interpreted by focusing on  an other aspect of the parameter estimation problem.
%From an intuitive perspective we expect  that increasing $k_1$ will result in the occurrence
%of more reactions during a given time interval, therefore more data will become  available
%for statistical inference and the variance of an unbiased estimator is expected to be decreased.
%This statement can be made precise again   by the pathwise FIM \VIZ{path_FIM} and the asymptotic
%normality of Maximum Likelihood Estimators (MLE), \cite{Kay:93, Wasserman:04}: the variance of the corresponding MLE  
% is the reciprocal of $k_1$ and it
%decreases when $k_1$ is increased. 
%Therefore, the minimum achievable variance of an
%unbiased estimator is also decreased.

\subsection*{The p53 Gene Model}
\label{p53:sec}
The p53 gene  plays a crucial role for effective tumor
suppression in humans as its universal inactivation in cancer cells suggests \cite{Prives:98,
Harris:05, Geva-Zatorsky:06}. 
The p53 gene is activated in response to DNA damage and gives rise to  a negative feedback
loop with the oncogene protein Mdm2. Models of negative feedback are capable of oscillatory
behavior with a phase shift between the gene concentrations. Here, we perform sensitivity
analysis to a simplified reaction network between three species,  p53, Mdm2-precursor and
Mdm2 introduced in \cite{Geva-Zatorsky:06}. The model  consists of five reactions and seven parameters
provided in Table~\ref{p53:reactions}. The nonlinear feedback regulator of p53 through Mdm2 
takes place in the second reaction while the remaining four reactions fall in the special class
where each reaction depends on one parameter.  Due to these mechanisms a nontrivial steady
state regime exists and can be characterized by random oscillations, see for instance
Figure~\ref{p53:species}. The proposed sensitivity methodology is directly applicable,
and the corresponding pathwise FIM, see \VIZ{FIM:log} and Figure~\ref{FIM:structure},
consists of 5 diagonal blocks with respective size $1\times1$, $3\times3$,
$1\times1$, $1\times1$, $1\times1$. Furthermore, the sensitivity analysis of this model has been
performed earlier  in \cite{Komorowski:11} based on a linear noise approximation. Here,
 we present a detailed comparison between the two sensitivity analysis methodologies, since the
 one proposed here does not involve any approximation of the stochastic network dynamics.

\begin{table}[!htb]
\begin{center}
\caption{The reaction table with $x$ corresponding to p53, $y_0$ to Mdm2-precursor while $y$
corresponds to Mdm2. The state of the reaction model is defined as ${\bf x}=[x,y_0,y]^T$
while the parameter vector is defined as $\theta=[b_x,a_x,a_k,k,b_y,a_0,a_y]^T$.}
\begin{tabular}{|c|l|l|l|} \hline
Event & Reaction & Rate & Rate's derivative\\ \hline \hline
$R_1$ & $\emptyset \rightarrow x$ & $a_1({\bf x}) = b_x$ & $\nabla_\theta a_1({\bf x}) = [1,0,0,0,0,0,0]^T$ \\ \hline
$R_2$ & $x \rightarrow \emptyset$ & $a_2({\bf x}) = a_x x + \frac{a_k y}{x+k} x$ & $\nabla_\theta a_2({\bf x}) = [0,x,xy/(x+k),-a_k xy/(x+k)^2,0,0,0]^T$ \\ \hline
$R_3$ & $x \rightarrow x+y_0$ & $a_3({\bf x}) = b_y x$ & $\nabla_\theta a_3({\bf x}) = [0,0,0,0,x,0,0]^T$ \\ \hline
$R_4$ & $y_0 \rightarrow y$ & $a_4({\bf x}) = a_0 y_0$ & $\nabla_\theta a_4({\bf x}) = [0,0,0,0,0,y_0,0]^T$ \\ \hline
$R_5$ & $y \rightarrow \emptyset$ & $a_5({\bf x}) = a_y y$ & $\nabla_\theta a_5({\bf x}) = [0,0,0,0,0,0,y]^T$ \\ \hline
\end{tabular}
\label{p53:reactions}
\end{center}
\end{table}

Figure~\ref{p53:species} shows the time-series of the species for the parameter
values in Table~\ref{p53:values}. Evidently, oscillatory behavior is observed at
this parameter regime, where  persistent random oscillations  occur, ranging between
high and low populations. On the other hand, the frequency
of the oscillations is less variable as it has been already reported both experimentally
and numerically \cite{Geva-Zatorsky:06}. Another interesting observation is that the
concentration of p53 species usually attains the lower bound of its admissible value
(populations cannot be negative) which results in stochastic effects far away from
Gaussianity, as can be readily seen also in Figure~\ref{p53:species}.

Proceeding, we denote by   $\theta = [b_x,a_x,a_k,k,b_y,a_0,a_y]^T$ the parameter vector.
The numerical estimator for RER as well as for the pathwise FIM in the logarithmic scale are
computed utilizing {\em Algorithm 1}. Logarithmic sensitivity analysis is preferred
because the range of the parameters values varies by orders of magnitude as can be seen in 
Table~\ref{p53:values}. The upper plot in Figure~\ref{p53:RER}
shows the RER as a function of time for various perturbations. Viewing 
RER as an observable, it is striking the speed of relaxation of the estimator.
Within two or three oscillation periods RER has been
converged to its value even though the three species have significant oscillations and stochasticity,
as Figure~\ref{p53:species} shows. A primary reason for the fast relaxation is the
numerical estimator of RER where the summation is over all reactions even though
only one reaction takes places at each jump (see \VIZ{FIM:num:approx:MP}).
Having the important property of fast convergence, global sensitivity analysis,
where not only a point of the parameter regime but also large subsets of the
parameter space, can be efficiently performed, \cite{Ludtke:08}.
The lower panel of Figure~\ref{p53:RER} shows the RER when only one of the
parameters are perturbed by +10\% or by -10\%. Additionally, the RER computed
from the FIM, utilizing \VIZ{GFIM},  is also provided. The FIM approximation
of RER is a second order approximation in terms of $|\epsilon|$, 
hence the computation of FIM is typically enough to fully resolve the local sensitivities of
a model. Evidently, the most sensitive parameters here are $b_x$ and $a_k$ while
the least sensitive parameters are $a_x$ and $k$.

\begin{table}[!htb]
\centering
\caption{Parameter values for the p53 model.}
\begin{tabular}{|c||c|c|c|c|c|c|c|c|} \hline
Parameter & $b_x$ & $a_x$ & $a_k$ & $k$ & $b_y$ & $a_0$ & $a_y$ \\ \hline
Value & 90 & 0.002 & 1.7 & 0.01 & 1.1 & 0.8 & 0.8 \\ \hline
\end{tabular}
\label{p53:values}
\end{table}

\begin{figure}[!htb]
\begin{center}
\includegraphics[width=\textwidth]{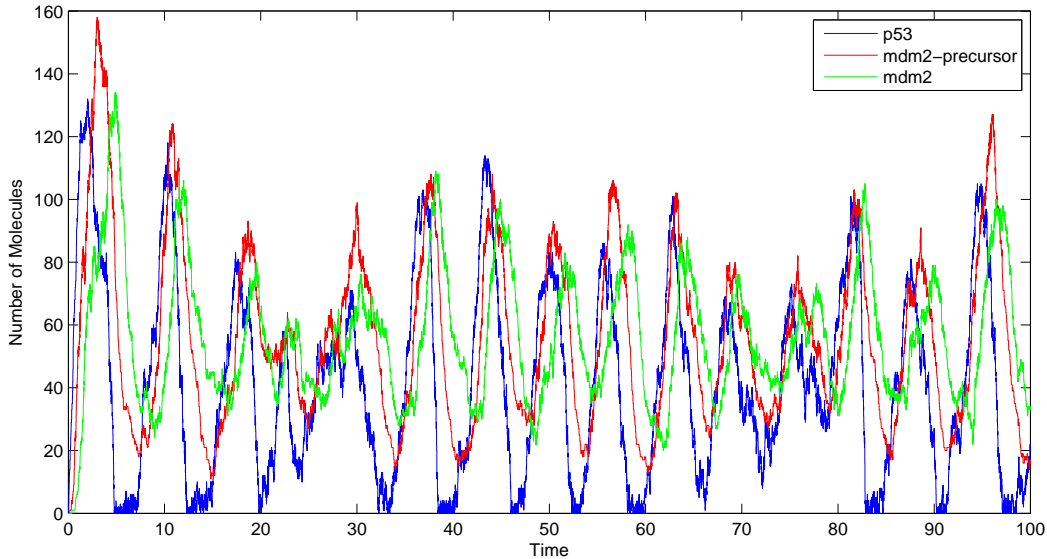}
\caption{Molecule concentration of p53, Mdm2-precursor and Mdm2. Concentration
oscillations as well as time delays (phase shifts) between the species are present due to the negative
feedback loop. Furthermore, the concentration of p53 periodically approaches  zero and
since negative concentrations are not allowed, the stochastic characteristics of p53
are far from Gaussian.}
\label{p53:species}
\end{center}
\end{figure}

\begin{figure}[!htb]
\begin{center}
\includegraphics[width=\textwidth]{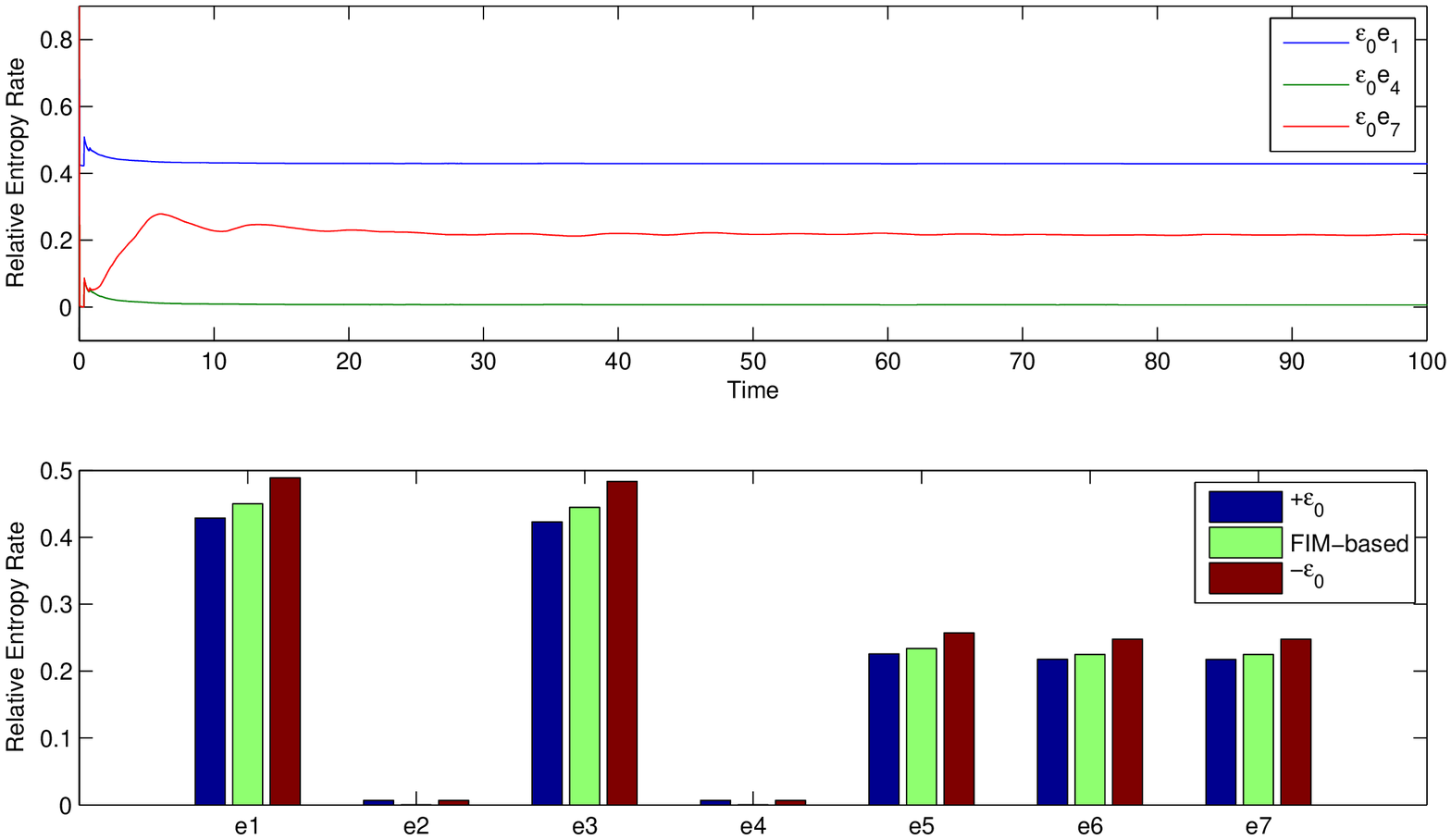}
\caption{Upper panel: RER in time for the parameter perturbation of $b_x$ (blue), $k$ (green) and
$a_y$ (red) by $+10\%$ (i.e., $\epsilon_0=0.1$). The relaxation time of the RER as an observable
is very fast. Lower panel: RER for various perturbation directions computed either directly (blue and
red bars) or based on FIM (green bars). Direction $e_k$ corresponds to perturbation of parameter
$\theta_k$.}
\label{p53:RER}
\end{center}
\end{figure}

\subsubsection*{Comparison to the LNA-based sensitivity approach}
In \cite{Komorowski:11}, the authors suggested a linear noise approximation (LNA) for
the stochastic evolution around the nonlinear mean-field equation, and based on
this approximation a system of ODEs is derived for the mean and the covariance
matrix of the approximation process. Since the noise of LNA is Gaussian, the mean
and the covariance matrix contains all necessary  information regarding the approximate
stochastic model. Then, the associated FIM is derived and based on it, the sensitivities
for each parameter are computed. Although there are  regimes where this approximation
is applicable (short times, high populations, systems with a single steady state, etc.),
for systems with nontrivial long-time dynamics, e.g. metastable, it is not correct as
large deviation arguments \cite{Doering:07} and explicit formulas for escape times
\cite{Hanggi:84} show. Similar issues  with non-gaussianity in the long-time
dynamics arise in stochastic systems with strongly intermittent (pulse-like) or random
oscillatory  behavior \cite{KMS}. In the p53 model considered in \cite{Komorowski:11}
which had the same parameter values as here, Figure~\ref{p53:species} reveals that
the time-series of the p53 populations persistently fluctuate between high and low
values, thus the LNA approximation may not be accurate at least 
when the concentration of the species is very low.

At  first pass, when the parameters are grouped into two classes depending on
their sensitivities, the two sensitivity approaches produce qualitatively similar results. Indeed, by visual
inspecting the lower plot of Figure~\ref{p53:RER} in the current publication and
Figure~3 in \cite{Komorowski:11}, the (more) sensitive parameters in both methods are
$b_x,b_y,a_k,a_0,a_y$ while the practically insensitive parameters are $a_x, k$. However,
upon closer inspection, the two methods produce different results. Figure~\ref{p53:FIM}
shows the proposed FIM (left) based on the exact (without any approximations)
pathwise relative entropy theory, as well the FIM proposed in \cite{Komorowski:11}
which is derived from the LNA of the reaction system. The results are completely
different and the proposed pathwise FIM is sparse as expected. 
A specific striking difference between the two sensitivity approaches  is that in our proposed
method the sensitivity of parameter $b_x$ is relatively high while in the LNA-based method
it is lower, see Figure~\ref{p53:FIM} (dark blue) and also compare   Figure~\ref{p53:RER}
of this publication  and  Figure~3 in \cite{Komorowski:11}. 

As a means of comparison  between the methods, we perturb $b_x$ as well as $b_y$ by 
the same amount and observe the Power Spectral Density (PSD), i.e., the square of the
absolute value of the Fourier transform of each species' time-series. Given the sustained
random oscillations observed in the p53 model, see Figure~\ref{p53:species}, the PSD is
a suitable observable since it identifies the dominant periodicities and corresponding 
amplitudes in stationary time-series, \cite{Kay:93}. Using  the Pinsker inequality
\VIZ{Pinsker:ineq} as a guideline, we expect that  the observable will not be sensitive
to the least sensitive directions of the pathwise FIM, therefore, we focus on the most
sensitive directions  of the FIM identified in Figure~\ref{p53:FIM}.
Figure~\ref{p53:pert} shows the averaged PSD
for the three species of the model for the unperturbed case (black lines),
the perturbation of $b_x$ only (blue lines) as well as the perturbation
of $b_y$ only (yellow lines). One hundred realizations were used for
the averaging procedure while the perturbation strength was +20\%.
It is evident that averaged PSD is more sensitive to perturbations of $b_x$
rather than to perturbations of $b_y$ as our sensitivity analysis
method predicted while the LNA-based method suggested the reverse order
of sensitivity.  Moreover, the $L_1$ norm between the unperturbed PSD and
and the $b_x$-perturbation is $8.56 \cdot 10^5$ while the $L_1$ norm between
the unperturbed PSD and and the $b_y$-perturbation is $4.32 \cdot 10^5$
which is about the half value.
An explanation of the  performance of the LNA-based sensitivity
analysis stems from the fact that the p53 species does not have Gaussian noise
when the population is close to zero, and which can  indeed occur frequently, see Figure~\ref{p53:species}
(blue line). Additionally, notice that both $b_x$ and $b_y$ affect the
concentration of p53 explicitly or implicitly through the associated reactions
thus their sensitivities are
heavily biased due to the wrong statistical approximation of the p53 species.

\begin{figure}[!htb]
\begin{center}
\includegraphics[width=\textwidth]{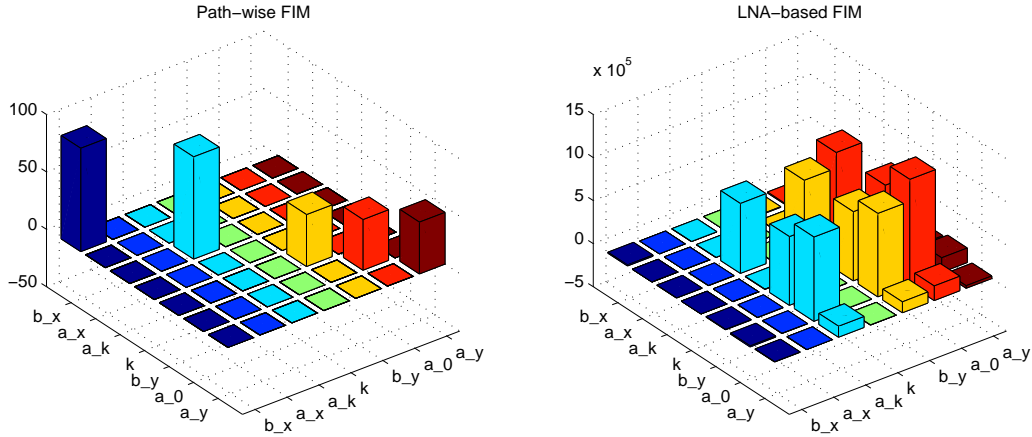}
\caption{The proposed pathwise FIM (left) based on RER as well as the (scaled) FIM based on
LNA computed from the StochSens package \cite{Komorowski:StochSens}. Evidently,
the proposed method uncouples the parameter correlations since most of the off-diagonal
elements are zero.}
\label{p53:FIM}
\end{center}
\end{figure}

\begin{figure}[!htb]
\begin{center}
\includegraphics[width=\textwidth]{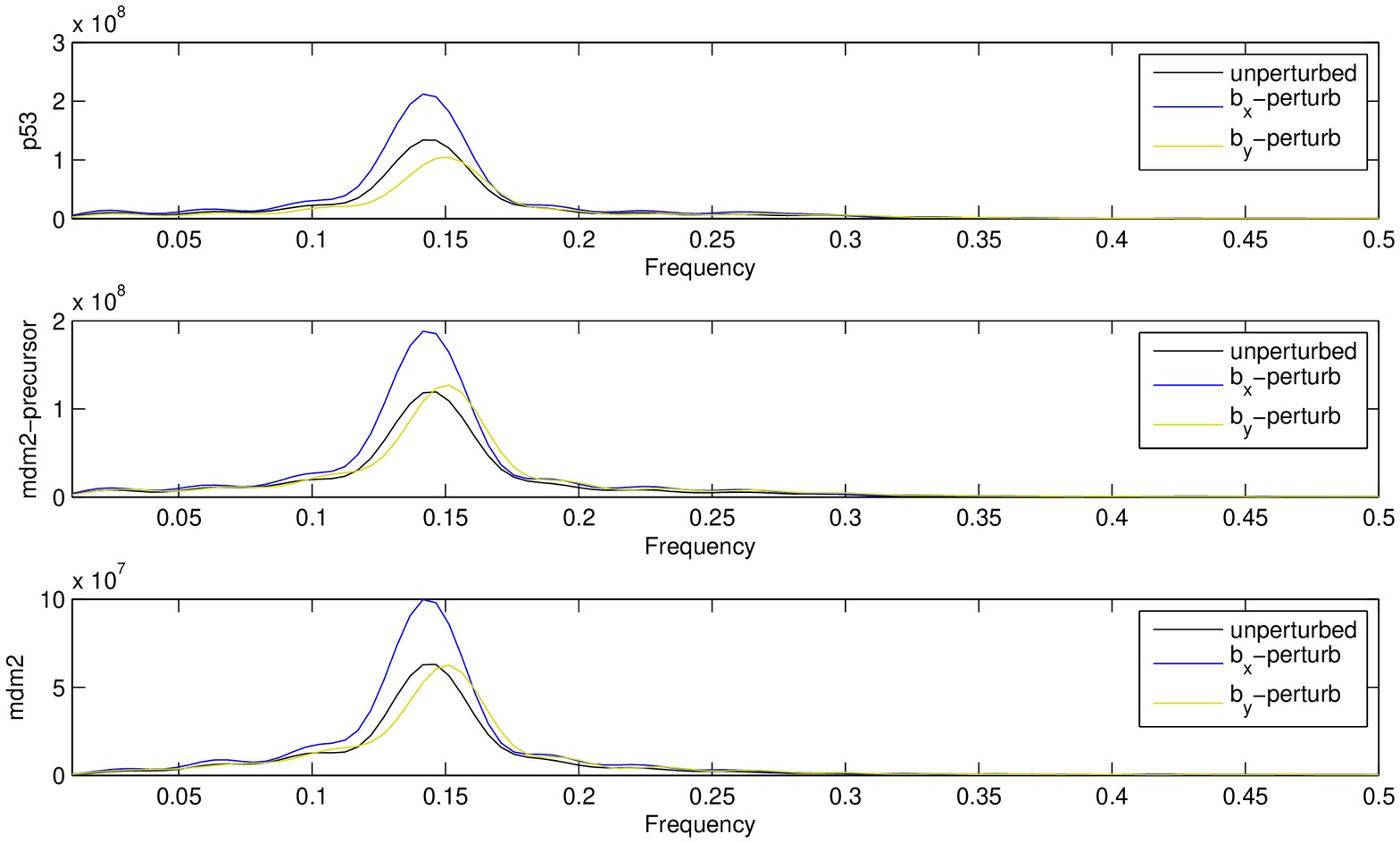}
\caption{Power Spectral Densities  (PSD) of the time-series of the species in the p53 model for the unperturbed parameter regime (black), when $b_x$
is perturbed by +20\% (blue) as well as when $b_y$ is perturbed by +20\% (yellow). The value and
the position of the prominent peak of the PSD are related with the amplitude and the frequency of the
species oscillations. The visual comparison between the averaged PSDs suggests that the spectral
properties are more sensitive to $b_x$ than to $b_y$.}
\label{p53:pert}
\end{center}
\end{figure}

\subsection*{Epidermal Growth Factor Receptor model}
\label{EGFR:sec}
The EGFR model is a well-studied system describing 
 signaling phenomena of (mammalian) cells \cite{Moghal:99, Hackel:99, Schoeberl:02}.
As its name suggests, EGFR regulates cell growth, survival, proliferation and differentiation
and plays a complex and crucial role during embryonic development and in tumor
progression \cite{Sibilia:98, Kim:99}. In this paper, we study 
the reaction network model  for  the dynamics of EGFR developed by Schoeberl
et al. \cite{Schoeberl:02} which consists of 94 species and 207 reactions.
Figure~\ref{EGFR:modules} presents the EGFR reaction network at an abstract level.
Initially, the extracellular binding of EGF with the EGF receptors induce receptor dimerization.
Then, two principal pathways, Shc-dependent and Sch-independend, are initiated
leading to activation of Ras-GTP. Subsequently  phosphorylation of MEK kinase through
the activation of Raf kinase occurs leading to the phosphorylation of ERK kinase
which regulates several proteins and nuclear transcription factors inside the cell.
The detailed graphical description of the reaction network can be found in the
Figures 1 \& 2 of supplementary information in \cite{Schoeberl:02}. For 
completeness, all the reactions along with their rates are provided
in the supplementary materials of this publication.
\begin{figure}[!htb]
\begin{center}
\includegraphics[width=\textwidth]{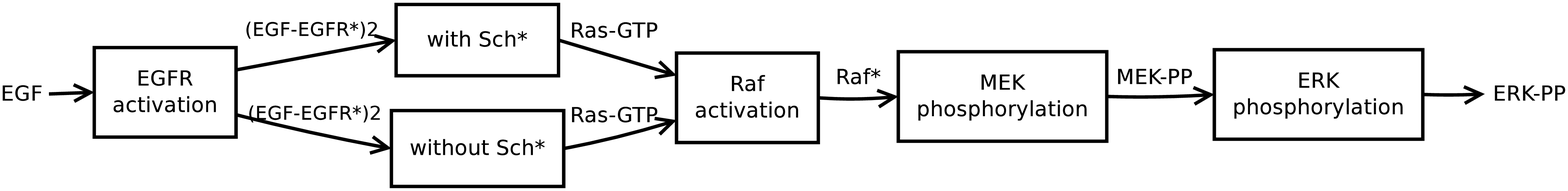}
\caption{Building blocks of the EGFR reaction network. Each module communicates
with the adjacent  modules through few species only. Additionally, with the exception of  the first module,
all the others are double, one external (i.e., outside the cell surface) and one
internal.}
\label{EGFR:modules}
\end{center}
\end{figure}
The propensity functions for the reactions $R_1,...,R_{97},R_{100},...,R_{207}$
of the EGFR network are written in the general form
\begin{equation}
a_j({\bf x}) = k_j \binom{{\bf x}_{A_j}}{\alpha_j} \binom{{\bf x}_{B_j}}{\beta_j},\ \ \ j=1,...,97,100,...,207
\end{equation}
with the exception of reaction pair $R_{98},R_{99}$ where their propensity functions are
governed by the Michaelis-Menten kinetics
\begin{equation}
a_j({\bf x}) = V_{\max} {\bf x}_{A_j} / (K_m+{\bf x}_{A_j}), \ \ \ j=98,99
\end{equation}
where ${\bf x}$ is the current state of the reaction system while $A_j$
corresponds to the reacting species. The parameter vector
contains all the reaction constants,
$$\theta=[k_1,...,k_{97},V_{\max}, K_m, k_{100},...,k_{207}]^T\, ,$$
 with all values  provided in the supplementary materials. Due to the specific values of the reaction
constants as well as the initial population of the species (see Table~\ref{EGFR:initial:data}),
the firing rates between reactions differ by many orders of magnitude giving rise to a highly stiff network.
Therefore, even though there are some  stochastic implementations, \cite{Chatterjee:05},
here for the purposes of RER and FIM calculations, we adopt the mean-field 
approximation  discussed in the accelerated estimators subsection. We solve
the derived system of ODEs with Matlab's routine ode15s and compute the FIM at the
steady state regime which corresponds to the time interval $[500,700]$. As Figure \ref{EGFR:k65}
suggests, the completion of the internalization process needs about 500 seconds. 
It should be noted here that even though the simulation of the EGFR is performed
utilizing a deterministic approximation model, the computed pathwise FIM has been
derived from the {\em stochastic} network, i.e., \VIZ{FIM:log}. This approximation is
expected to be valid in the sense of  \VIZ{approx:MFE} due to the large populations
considered here. Overall, the computed FIM is a sparse matrix and measures efficiently
the sensitivities of the stochastic model in a gradient-free manner.

\begin{table}[!htb]
\centering
\caption{Initial population of the species for the EGFR network.}
\begin{tabular}{|c|c|c|c|c|c|c|} \hline
EGF & EGFR & GAP & Grb2 & Sos & Ras-GDP & Shc  \\ \hline
4.98e10 & 5e4 & 1.2e4 & 5.1e4 & 6.63e4 & 1.14e7 & 1.01e6  \\ \hline\hline
Raf & Phosphatase 1 & Phosphatase 2 & Phosphatase 3 & MEK & ERK & Prot \\ \hline
4e4 & 4e4 & 4e4 & 1e6 & 2.2e7 & 2.1e7 & 8.1e4 \\ \hline
\end{tabular}
\label{EGFR:initial:data}
\end{table}

The upper plot of Figure~\ref{EGFR:FIM:fig} shows the diagonal elements of the FIM in
descending order computed at the steady state regime. We report our results in the format
of Figure~\ref{EGFR:FIM:fig} in order to be able to accommodate the large number
of parameters in the model. The $k$-th diagonal
element of the FIM corresponds to RER where the perturbation takes place only to the
$k$-th parameter (see \VIZ{GFIM}). Figure~\ref{EGFR:FIM:fig} (upper plot) in conjunction
with Table~1 of the supplementary file EGFR\_table.pdf fully describe the (local) sensitivities
of the reaction network. Table~1 in EGFR\_table.pdf presents the reaction constants ordered
from the most sensitive to the least sensitive parameter.
Moreover, the FIM is diagonal --except a small $2\times2$ block associated
with the Michaelis-Menten reactions-- therefore the diagonal elements correspond to the eigenvalues
of the FIM. The sensitivity analysis depicted in Figure~\ref{EGFR:FIM:fig}, demonstrates
that most model parameters allow for a vast range of perturbations without affecting the
dynamics. Furthermore, this robustness
to variations in most parameters was also reported in the original, fully deterministic EGFR model
in \cite{Schoeberl:02}. This is a feature shared by  many multi-parameter models in systems
biology and which is known as "sloppiness", \cite{Gutenkunst:07}. Our methodology  can
easily demonstrate such  properties in stochastic dynamics, as we can readily see in Figure~\ref{EGFR:FIM:fig},
even if the models include a large number of parameters.

The previous  discussion refers to the analysis of the EGFR model to the steady state regime.
On the other hand, EGFR is a signaling model whose transient regime, in addition to
the steady state, is of great interest. As discussed in Remark 3,
we can justify the application of  the RER and FIM sensitivity analysis  in the transient regime.
Therefore, we compute the proposed FIM at the time interval $[0,10]$, using \VIZ{spe:case:num:FIM}.
The lower plot of Figure~\ref{EGFR:FIM:fig} shows the diagonal elements of the pathwise FIM in
the transient regime while keeping the ordering of the parameters unchanged from the
upper, steady state  plot. The parameter sensitivity ordering is
completely different meaning that the sensitivities are time-dependent in the transient regime.
For instance, the most sensitive parameters in the stationary regime correspond to the final
products of the reaction network, however, in the time interval $[0,10]$ these species have not
been produced yet resulting to insensitive reaction constants. In terms of parameter identification and estimation,
the time-dependent sensitivities imply that in order to extract the maximum information
content from the experimental data, we have to estimate the parameters drawing  samples
from different time intervals. These time intervals should be defined based on the respective
sensitivity indices and selected in order to maximize the identifiability for each set of parameters.

\begin{figure}[!htb]
\begin{center}
\includegraphics[width=\textwidth]{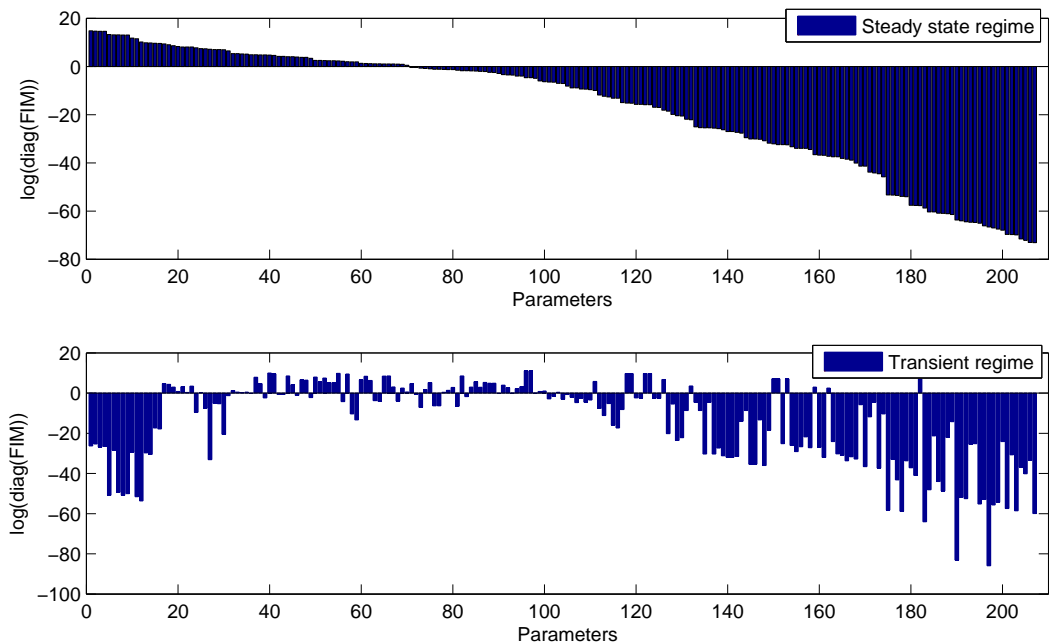}
\caption{Diagonal elements of the FIM computed at the steady state regime (upper plot)
and at the transient regime (lower plot). Note the changes in sensitivity and consequently
the parameter identifiability.  The parameter sensitivities differ by orders of magnitude.}
\label{EGFR:FIM:fig}
\end{center}
\end{figure}

Finally, the Pinsker inequality \VIZ{Pinsker:ineq} suggests that insensitive parameters can
be perturbed,  even significantly, without affecting  species concentrations or  other observable.
As an illustration of this fact, we present  in Figure \ref{EGFR:k65} the concentrations of
various critical species of the EGFR model when the 140-th ($k_{65}$) most sensitive
parameter is perturbed (see Table~1 in supplementary materials). The rate constant $k_{65}$
corresponds to a reaction of the Shc-dependent pathway module. Solid blue lines correspond
to the unperturbed parameter case while the dashed red lines correspond to the perturbed
case where the perturbation is a multiplication by a factor of ten of $k_{65}$.
We  present the total number of (EGF-EGFR*)2 binding species without
Sch* (top, left panel) and with Sch* (top, middle panel) as well as Ras-GTP (top, right
panel), total activated Raf or total Raf* (low, left panel), doubly phosphorated
MEK or MEK-PP (low, middle panel) and doubly phosphorated ERK or ERK-PP.
These species are important for the understanding of the system since the different
modules of the EGFR reaction network communicate  through them
(see Figure~\ref{EGFR:modules}). It is evident from Figure~\ref{EGFR:k65} that the
various species concentrations remain unchanged to perturbations of the insensitive
parameter $k_{65}$ as it was expected from \VIZ{Pinsker:ineq}. Moreover, we notice that although the average populations
become large, which implies that the maximum norm in \VIZ{Pinsker:ineq}  is also large, we still obtained 
robust results regarding  $k_{65}$'s parameter  sensitivity.

\begin{figure}[!htb]
\begin{center}
\includegraphics[width=\textwidth]{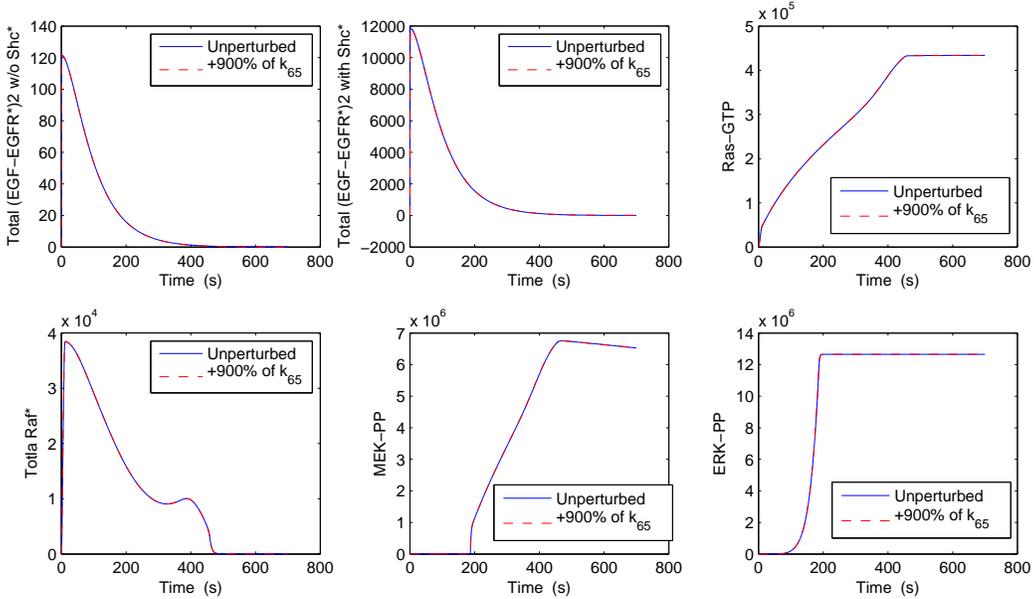}
\caption{Time-dependent concentration of various species of the EGFR network either for the
unperturbed parameter vector (solid blue lines) or for the perturbed one (dashed red lines).
The 140-th most sensitive parameter ($k_{65}$) is ten-fold increased and the species
concentrations are not affected. For the least sensitive parameters such as $k_{65}$,
we rigorously know from the  Pinsker inequality \VIZ{Pinsker:ineq} that they should not alter
the concentration values or any other observable even when they are heavily perturbed.}
\label{EGFR:k65}
\end{center}
\end{figure}

\section*{Conclusions}
\label{concl:sec}
In this paper, we applied and extended a recently proposed parametric sensitivity analysis
methodology to complex stochastic reaction networks. This sensitivity analysis approach
is based  on the quantification of information loss along different parameter perturbations
between time-series distributions. This is achieved by employing the
Relative Entropy Rate, which is directly  computable from the propensity functions.
A key aspect of the method is that we can derive rigorously an associated Fisher
Information Matrix on path-space, which in turn constitutes a gradient-free approach for 
 parametric sensitivity analysis; as such it provides a significant  advantage in
stochastic systems with a large number of parameters.
We  demonstrated that  the structure of the pathwise FIM revealed hidden, parameter
interdependencies between the reactions. The block-diagonal structure
of the FIM highlighted the sparsity of the matrix which resulted in further improvements 
in  the computational  efficiency of the proposed method. Therefore, parametric sensitivity analysis
for high-dimensional stochastic reaction systems becomes  tractable since it is well-known
that in high dimensional stochastic systems  sensitivity analysis techniques can involve
estimators of very high variance, e.g. in finite difference methods and their recently
proposed variants, which can present an overwhelming computational cost.
Additionally, we proposed the use of  multiscale numerical approximations of stochastic
reaction networks in order to derive efficient statistical estimators for the FIM and implemented 
one such approximation (mean-field) in  a high-dimensional system.

The proposed pathwise sensitivity analysis method is tested and validated on three biological
systems: (a) a simple protein production/degradation model where explicit solutions are available,
(b) the p53 reaction network where quasi-steady stochastic oscillations of the
concentrations are observed and where multiscale stochastic approximations break down due to
the persistent oscillations  between low and high populations, and (c)  a stochastic EGFR
model  which is an example of a high-dimensional  reaction
network with more than 200 reactions and a corresponding number of parameters.
In the EGFR reaction network, we combined the proposed pathwise FIM which has
been derived from the stochastic network and the mean-field approximation which is
used for the efficient estimation of the pathwise FIM. Moreover, our earlier rigorous analysis
for the steady state regime \cite{Pantazis:Kats:13} suggests suitable extensions in
the transient regime which were  tested and validated for the EGFR model.
We will present the full rigorous theory in an upcoming publication.

Finally,  we note that the relation between RER and various observables is not
straightforward. However, we note that the path
distribution contains all information regarding the process including the steady state  and
all time-dependent observables: practically, our proposed sensitivity analysis represents a
ÒconservativeÓ sensitivity estimate in the sense that insensitive directions for the relative
entropy on path-space, will yield insensitive directions for every observable. This latter
statement can be justified  mathematically through  the Pinsker inequality \VIZ{Pinsker:ineq}
which was tested in the examples considered here.
Based on these observations,  the proposed sensitivity analysis  methods can be 
deployed  in complementary fashion with existing sensitivity analysis tools, as it can be used
to narrow down the most sensitive directions in a system.

\section*{Competing Interests}
The authors declare that they have no competing interests.

\section*{Author's Contributions}
MAK and YP conceived the proposed sensitivity analysis methodology. 
YP conducted the numerical experiments. MAK and
DGV designed and supervised the conducted research.  All authors contributed to
the preparation of the manuscript.

\section*{Additional Material}
\begin{itemize}
\item \begin{verbatim}derivation_RER_FIM.pdf:\end{verbatim} The detailed derivation of relative entropy rate and
the associated Fisher information matrix.
\begin{verbatim}derivation_birthDeath_FIMs.pdf:\end{verbatim} The calculation of equilibrium and pathwise
FIMs for the protein production/degradation model.
\item \begin{verbatim}EGFR.txt:\end{verbatim} This file contains in plain text the reactions and the reaction constants
of the EGFR model.
\item \begin{verbatim}EGFR_table.pdf:\end{verbatim} The ordering of the
parameter sensitivities for the EGFR model.
\end{itemize}

\section*{Acknowledgements}
The work of MAK and YP   was supported in part by the Office of
Advanced Scientific Computing Research, U.S. Department
of Energy under Contract No. DE-SC0002339. 
The work of DGV was supported in part by the Office of
Advanced Scientific Computing Research, U.S. Department
of Energy under Contract No. DE-FG02-05ER25702.
The work of MAK was also supported in part by   the European Union
(European Social Fund) and Greece (National Strategic Reference Framework),
under the THALES Program, grant AMOSICSS.

{\ifthenelse{\boolean{publ}}{\footnotesize}{\small}
\bibliographystyle{bmc_article}  % Style BST file
\bibliography{biblio}
}

\ifthenelse{\boolean{publ}}{\end{multicols}}{}

\end{bmcformat}
\end{document}